\pdfoutput=1

\documentclass[11pt]{article}

\usepackage[final]{acl}

\usepackage{times}
\usepackage{latexsym}

\usepackage[T1]{fontenc}

\usepackage[utf8]{inputenc}

\usepackage{microtype}

\usepackage{inconsolata}

\usepackage{amsmath}

\usepackage{graphicx} 
\usepackage[capitalise]{cleveref} 
\usepackage{booktabs} 

\usepackage{array} 

\usepackage{amssymb}
\usepackage{multirow}

\newcommand{\recall}[1]{{\em recall@#1}}
\newcommand{\ourdataset}{D-MERIT}

\title{Evaluating \ourdataset{} of Partial-annotation on Information Retrieval}

\author{Royi Rassin\textsuperscript{\normalfont1$\thanks{~~This project was done during an internship.}$,2} \, Yaron Fairstein\textsuperscript{\normalfont 1}\, Oren Kalinsky\textsuperscript{\normalfont 1}\, Guy Kushilevitz\textsuperscript{\normalfont 1}\\ \textbf{Nachshon Cohen}\textsuperscript{\normalfont 1}\ \textbf{Alexander Libov}\textsuperscript{\normalfont 1}\, \textbf{Yoav Goldberg}\textsuperscript{\normalfont2,3}\\
\textsuperscript{1}Amazon Research \,\textsuperscript{2}Bar-Ilan University \, \textsuperscript{3}Allen Institute for AI \\\\ 
    {\tt\{\href{mailto:rassinroyi@gmail.com}{rassinroyi},
    {\href{mailto:Yyfairstein@gmail.com}{yyfairstein},
    {\href{mailto:orenkalinsky@gmail.com}{orenkalinsky},
    {\href{mailto:yoav.goldberg@gmail.com}{yoav.goldberg}\}@gmail.com}}}}\\
    {\tt\{\href{mailto:guyk@amazon.com}{guyk}},
    {\href{mailto:nachshon@amazon.com}{nachshon}},
    {\href{mailto:alibov@amazon.com}{alibov}\}@amazon.com}}
    
\begin{document}
\maketitle

\begin{abstract}
Retrieval models are often evaluated on partially-annotated datasets. Each query is mapped to a few relevant texts and the remaining corpus is assumed to be irrelevant. As a result, models that successfully retrieve falsely labeled negatives are punished in evaluation. Unfortunately, completely annotating all texts for every query is not resource efficient. In this work, we show that using partially-annotated datasets in evaluation can paint a distorted picture. We curate \ourdataset{}, a passage retrieval evaluation set from Wikipedia, aspiring to contain \emph{all} relevant passages for each query. Queries describe a group (e.g., ``journals about linguistics'') and relevant passages are evidence that entities belong to the group (e.g., a passage indicating that \textit{Language} is a journal about linguistics). We show that evaluating on a dataset containing annotations for only a subset of the relevant passages might result in misleading ranking of the retrieval systems and that as more relevant texts are included in the evaluation set, the rankings converge. We propose our dataset as a resource for evaluation and our study as a recommendation for balance between resource-efficiency and reliable evaluation when annotating evaluation sets for text retrieval. Our dataset can be downloaded from \tt\href{https://D-MERIT.github.io}{\texttt{https://D-MERIT.github.io}}.
\end{abstract}

\section{Introduction}

\begin{figure}
    \centering
    \includegraphics[width=1\columnwidth]{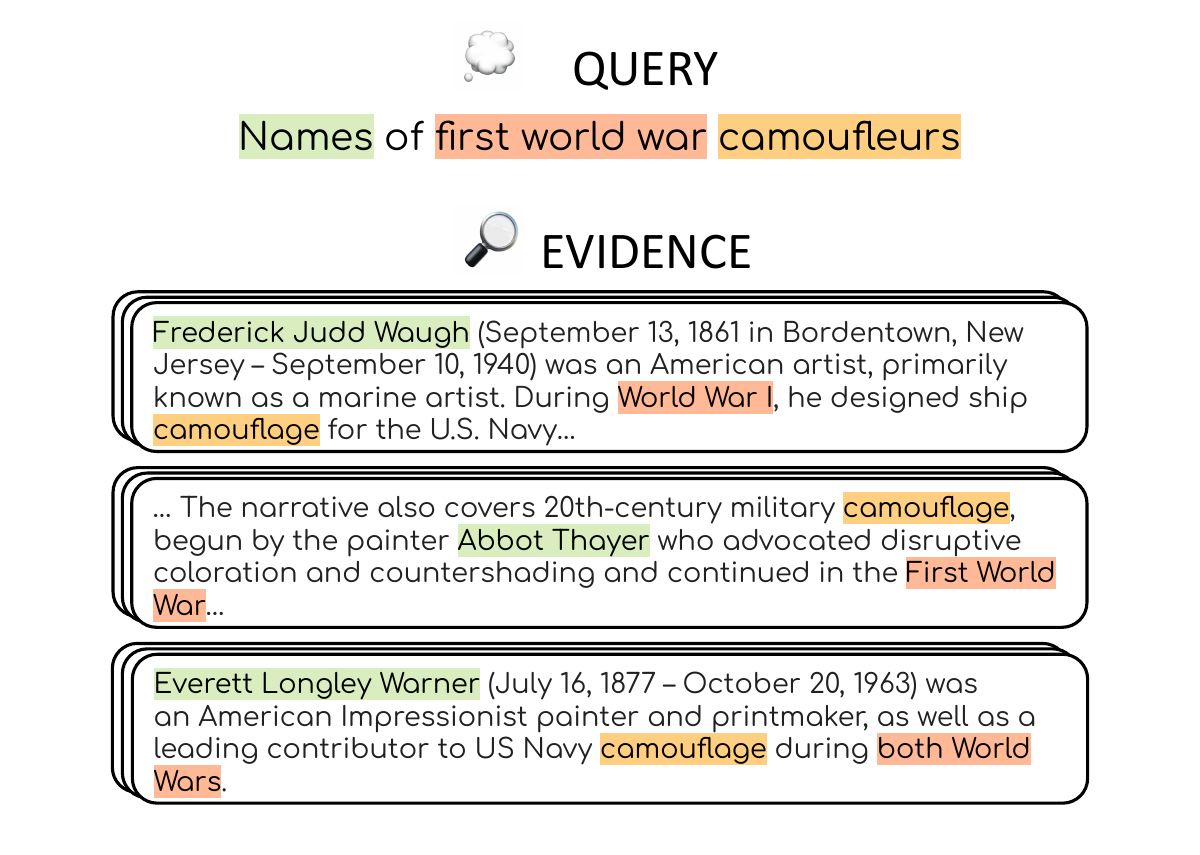}
    \caption{Demonstrating the evidence retrieval task described in \cref{sec:evidence_retrieval_task}. The query is ``Names of first world war camoufleurs''. Highlighted text corresponds to the query requirements: names (green), ``First World War'' (red), and ``camouflage'' (orange). A passage must match all requirements to be considered as evidence.}
\label{fig:task_figure}
\end{figure}

Passage retrieval, the task of retrieving relevant passages for a given query from a large corpus, is a traditional IR task \cite{kaszkiel1997passage,callan1994passage,zobel1995efficient}. Within NLP, it has many applications, such as Open-Domain Question-Answering (ODQA) \cite{dpr,zhu2021retrieving,mavi2022survey,rogers2023qa} and fact verification \cite{bekoulis2021review,murayama2021dataset,vallayil2023explainability}. 

Recently, the task has experienced a renaissance due to the modern retrieval-augmented-generation setup leveraging LLMs (aka ``RAG'') \cite{rag_paper,cai2022recent,li2022survey}. In all of those cases, retrieval makes for a crucial component of the system \cite{cai2022recent, ram2023context}. 

It is common practice, and often essential to evaluate the retriever component separately from the full system. This is done by using large-scale data resources that map queries to relevant passages.\footnote{Relevancy is defined according to the task in hand. In this work, we adopt the definition of TREC \citep{trec2019}, a popular retrieval research challenge.} The vast majority of available datasets are only partially-annotated; a query is mapped to a single (or a few) relevant passages and all other passages are assumed to be irrelevant \citep{msmarco, natural_questions}, leading to many passages falsely labeled as negatives in the dataset. This practice has long been contested \citep{how_reliable_are_the_results_of_large_scale_info_retrieval_experiments, retrieval_with_incomplete_info, trec2019, survivorship_bias_msmarco}, yet due to the massive size of modern corpora, exhaustively annotating all passages for every query is highly impractical. As an example, MS-MARCO \citep{msmarco} consists of \textasciitilde 1M  queries and \textasciitilde 8.8M passages, which amounts to \textasciitilde 8.8 \emph{trillion} annotations. 

Evaluating retrieval solutions using a partially-annotated dataset is obviously not ideal. A system retrieving a non-annotated relevant passage rather than an annotated one is unjustly penalized. Some work has been done on metrics and methods attempting to deal with this issue \citep{retrieval_with_incomplete_info, infAP, one_shot_labeling_relevance}. However, the common practice is still using vanilla metrics (e.g. $MRR$, $Recall$), and the impact of partial annotation during evaluation using these metrics is still unclear. Does the ranking of systems change? Do the inaccurate scores falsely crown the wrong systems as the SOTAs? Moreover, we wonder how many relevant passages are needed in order to sufficiently reduce the error and correctly rank systems.

In this work, we propose \textbf{\ourdataset{}}; \emph{\textbf{D}ataset for \textbf{M}ulti-\textbf{E}vidence \textbf{R}etr\textbf{i}eval \textbf{T}esting}, an evaluation set for retrieval systems, \emph{striving} to pair each query to \emph{all} of its relevant passages. In our setting, relevant passages are evidence that some entity belongs to a group described in the query. While we use it to explore the consequences of having an evaluation dataset with only a few relevant passages annotated, \ourdataset{} is also highly suitable for use in high-recall settings, where the task is to retrieve as many relevant texts as possible for a given query, as it contains almost all relevant passages available in the corpus for each query.

We first show that evaluation of systems with the common single-relevant setup (for each query, annotate passages until a single relevant passage is found) is sensitive to the way in which passages were selected during annotation. As a result, different selections lead to different rankings of systems. 
However, we observe that when a system very significantly outperforms another, representing a seminal improvement or breakthrough, the single-relevant setup is likely to provide accurate rankings. 
Then, we mimic partially-annotated setups, gradually adding annotated relevant passages to queries, hence reducing the number of falsely labeled negatives in the data. Our findings reveal that in order to reliably evaluate retrieval systems that are reasonably close in performance, a significant portion of relevant passages must be found. This is substantial because it implies that when evaluating using partially-annotated datasets, some system might \emph{seem} better-performing than another, while in fact, the opposite is true.
To summarize, our contributions are as follows:
\begin{itemize}
    \item \ourdataset{}: A publicly available passage retrieval evaluation set, aspiring to contain all relevant passages per query.
    \item A study on the consequences of leaving too many falsely labeled negatives in evaluation sets.
    \item Recommendations for a balance between resource-efficiency and reliable evaluation when annotating retrieval datasets.
\end{itemize}

\section{\ourdataset{}}

\subsection{Desiderata} To observe the impact of having falsely labeled negatives in an evaluation set, we need to have a dataset where the falsely labeled negatives are marked as such. This calls for a completely-annotated dataset, that will allow us to reliably evaluate systems' performance, as well as examine the effects of partial-annotation. To accentuate the gap between partial and full annotation, queries in the dataset should be mapped to many relevant passages.
We are set to try to identify \textit{all} relevant passages for each query, but annotating all passages for each query is unrealistic. Therefore, we desire a framework that offers inherent mappings between queries and high quality candidate passages. To push our method towards exhaustiveness, our automatic approach to candidate collection needs to lean towards recall, followed by an automatic filtering stage.

\subsection{Task Definition} \label{sec:evidence_retrieval_task}
\paragraph{Evidence Retrieval.} We choose evidence retrieval as our task as it naturally complements our need to collect queries with numerous relevant passages. 
In this task, passages are considered relevant if they contain text that can be seen as evidence that some answer satisfies the query. Previous work considering this task did not collect more than a single evidence \citep{quest, qampari} or did not aspire to be completely-annotated \citep{romqa}. Instead, they map queries to answers, and collect evidence for each answer from a single document. Our goal is to map a query to \emph{all} evidence in the corpus, without the limitation of a single document.

\paragraph{Our setup.} In our setup, that can be seen as an extension of the single-evidence setup in \citep{quest} to an all-evidence one, a query describes a group of entities and relevant passages are evidence that an entity is a member of the group. The task is then, given a query representing some group, to retrieve all texts stating that some entity is a part of this group. 
For instance, \cref{fig:task_figure} shows evidence for the query ``names of first World War camoufleurs''. The first passage confirms ``Fredrick Judd Waugh'' is an entity that belongs to the group of World War~1 camoufleurs. More concretely, each query lists constraints, and an evidence would associate an entity with all of them.\footnote{The queries in our setup are somewhat reminiscent to the intersection queries in \citep{quest}, where a query makes for a list of requirements.} In the example above, a query describes the group of all World War 1 camoufleurs, an evidence would then need to indicate an entity (1) took part in World War 1; (2) was a camoufleur. For example, the second passage in \cref{fig:task_figure} states ``Abbot Thayer'' advocated for coloration and countershading camouflage during World War 1, which satisfies these requirements.

\subsection{Dataset Curation}
We adopt the Wikipedia framework\footnote{The Wikidump is from July 1st, 2023.}, which allows us to take advantage of the Wikidata structure \citep{wikidata} to extract groups and their corresponding members. We use the Wikipedia link network to obtain mappings between an article and all other articles referencing it. Our curation process involves three stages: (1) collecting queries and \emph{candidates} -- all passages with high likelihood of containing evidence (\cref{sec:candidate_collection}); (2) automatic annotation of candidate passages (\cref{sec:evidence_identification}); (3) generating natural language queries (\cref{sec:nl_query_generation}). 

\subsubsection{Corpus} Our corpus is limited to the introduction section of Wikipedia articles. Without limiting our collection process to a specific section, the number of annotations per article would have multiplied by \textasciitilde 5, which would have made the annotation process significantly more expensive. We opted to focus on the introduction section, because it is a section that is consistent across most articles, and it is intuitive that many evidence lie there. In total, our corpus is comprised of $6,477,139$ passages.

\begin{table*}[t]
\centering
\scalebox{0.80}{
\begin{tabular}{>{\raggedright\arraybackslash}p{3cm}|>{\centering\arraybackslash}p{2cm}|>{\centering\arraybackslash}p{2cm}|>{\raggedright\arraybackslash}p{10cm}}
\textbf{Query} & \textbf{Member} & \textbf{Candidate} & \textbf{Evidence} \\
\hline
names of Indian Marathi romance films & Sairat & Jeur & Jeur is a village in the Karmala taluka of Solapur district in Maharashtra state, India.
\textbf{Sairat}, the controversial and highest-grossing \textbf{Marathi film} of all time based on the theme of \textbf{forbidden love} was set and shot in Jeur village.\\
\hline
names of National Wildlife Refuges in West Virginia & Ohio River Islands National Wildlife Refuge & Mill Creek Island &
Mill Creek Island is a bar island on the Ohio River in Tyler County, \textbf{West Virginia}. The island lies upstream from Grandview Island and the towns of New Matamoras, Ohio and Friendly, West Virginia. It takes its name from Mill Creek, which empties into the Ohio River from the Ohio side in its vicinity. \textbf{Mill Creek Island is protected as part of the Ohio River Islands National Wildlife Refuge}.\\
\hline
Names of players on 1992 US Olympic ice hockey team & Dave Tretowicz & Dave Tretowicz & \textbf{Dave Tretowicz} (born March 15, 1969) is an \textbf{American former professional ice hockey player}. In 1988, he was drafted in the NHL by the Calgary Flames. \textbf{He competed in the men's tournament at the 1992 Winter Olympics}. \\
\end{tabular}}
\caption{Examples of records in our dataset. \textbf{Query} is the generated natural-language query describing a group. \textbf{Member} is an entity that belongs to the group described by the query. \textbf{Candidate} is the Wikipedia article from which the evidence is taken from. \textbf{Evidence} is a passage indicating the member's association with the group.}
\label{tab:evidence_examples}
\end{table*}

\subsubsection{Query and Candidate Collection} \label{sec:candidate_collection}
\paragraph{Extracting list members.} The collection process begins by scanning articles prefixed with ``list of’’ for tables using the Wikidata format. We extract columns with ``name'' in their title, as these are most likely to describe entities. Each such column is extracted separately and makes for a set of members. Columns containing empty values or values without a dedicated Wiki article are discarded.

\paragraph{Collecting candidates} We employ the "What Links Here" feature from Wikidata. This tool provides a list of all articles that reference a specific article (and its aliases). The reference count of an article can vary significantly, even for members of the same list. For example, ``Shogi'' has over 600 references, while ``Machi Koro'' only has 9. Both appear in the group ``Japanese board games''. To manage this disparity and keep the candidate count feasible, we discard columns containing an article with more than $10K$ references.

\subsubsection{Evidence Identification} \label{sec:evidence_identification}
To complete the dataset construction, we need to sift through the collected candidates. Human evaluation would have been the most reliable route, however, it does not scale. We thus turn to the current state-of-the-art large language model for automatic filtering, and show it nears human judgement.

\paragraph{Automatic identification.} 
We use \texttt{GPT-4}\footnote{We used \texttt{GPT-4-1106-preview}. Future references to \texttt{GPT-4} refer to this version.} 
to filter
$\sim250K$ passages across
$\sim2.5K$ queries. 
Each prompt consists of a passage paired with a query embedded in our definition of relevance, asking the model to judge for relevance. To ensure each query is meaningful in number of evidence, queries with less than five evidence were discarded. For technical details, see \cref{app:automatic_filtering}.

\subsection{Evaluation of Construction Process} \label{sec:eval_of_dataset_construction}
In order for \ourdataset{} to contain a significant portion of the positives for each query, some assumptions need to hold. First, Wikipedia list pages need to be exhaustive.\footnote{Note that we only need the list to be exhaustive with respect to the corpus, i.e. if some set member is not in the list but is also not mentioned in Wikipedia introductions, it will not hinder the exhaustiveness of our collection method.} This is a common assumption also taken by \cite{qampari} and \cite{quest}. Our dataset construction method also relies on the accuracy of Wikipedia's linking network. This is a limitation of the method (and is therefore mentioned in the limitations section). Herein, we want to show these assumptions do not meaningfully degrade the quality of the dataset. To this end, we approximate \ourdataset{}'s completeness and soundness by evaluating the candidate collection process -- if we have missed a meaningful number of evidence during candidate collection. To complete the evaluation of \ourdataset{}'s quality, we also evaluate our automatic identification model, \texttt{GPT-4}, to confirm it reliably identifies the vast majority of evidence without adding much false positives.

\paragraph{Evaluation tasks.} We turn to Amazon Mechanical Turk (AMT) for sourcing human raters. For the candidate collection evaluation, a human rater is provided with a passage and a prompt containing the query, and is requested to mark whether the passage is evidence or not. In the task designed to gauge the quality of the automatic identification, in addition to the passage and prompt, the annotation of \texttt{GPT-4} is also provided. The rater is then requested to judge the correctness of the annotation. Since judging relevance can be subtle\footnote{Consider row 2 in \cref{tab:evidence_examples}, where the passage does not explicitly say that “Ohio River Islands National Wildlife Refuge” is in “West Virginia”. Instead, it says that “Mill Creek Island”, which is in “West Virginia”, is part of the “Ohio River Islands National Wildlife Refuge”.}, we make a decision to judge the correctness of annotations, instead of to annotate and compare results to \texttt{GPT-4}. This encourages the rater to consider the annotation's perspective and allows tolerance toward borderline cases. The selection and conditioning process of human raters is detailed in \cref{app:further_details_dataset_creation}.

\paragraph{Exhaustiveness of candidate collection.}To ensure our collection process is nearly exhaustive, we need another evidence collection process, independent of ours. 
We thus adopt the popular TREC approach \citep{trec2019}, where a number of systems retrieve the top-$k$ passages given a query, and are then unified to a single set of passages to be judged for relevancy. We use $12$ different systems, described in \cref{sec:experiment_setup}.
As for the pool depth, we select $k=20$ to match our experimental study. Several works researched the relation between pool depth and the completeness of TREC evaluations \cite{bias_of_pooling, effect_of_pool_depth_trec_eval, effect_of_pooling_on_ir_metrics} raising concerns regarding reliability of the shallow pool depth commonly used (the typical TREC setup uses a $k=10$ depth), hence we also extrapolate the results of this evaluation to a $k=100$ pool depth.

We select $23$ random queries from \ourdataset{}, and use the TREC approach to retrieve $2,329$ unique passages. Since we are looking for relevant passages that we missed, we discard unique passages that were already annotated by our process ($311$ such cases, all relevant) and are left with $2,018$ passages. We ask human raters to mark the remaining passages for relevance and find \emph{only} $35$ new evidence. 
In total, the TREC process finds $346$ relevant passages, $311$ of which were found by our process too. To put this in context, for the same $23$ queries, our process finds $990$ relevant passages. We note that while our method retrieves many more evidence, it is tailor-made to the Wikidata format, while the method from TREC can be applied to any corpus.
To further attest to the exhaustiveness of our approach, we extrapolate the analysis to $k=100$, and estimate the number of identified evidence to increase to $638$, with only $60$ new evidence. A more profound discussion of TREC's coverage, including details on the extrapolation process, can be viewed in \cref{app:trec_coverage_simulation}.

To summarize, the TREC process, with a pool depth of $k=20$, finds $346$ positives and requires $2,329$ annotations ($\sim14.9\%$ positives in the pool). Our method finds $990$ positives, requiring $3,206$ annotations ($\sim30\%$ positives in the pool). The TREC process adds only $\sim3.5\%$ new positives to our method. When TREC is extrapolated to a pool depth of $k=100$, \ourdataset{} still has a high (estimated) coverage of $94.5\%$ of identified evidence.

\paragraph{Comparing automatic to manual identification.}
To verify \texttt{GPT-4} is comparable to manual identification, we collect a random sample of $1,300$ (query, passage) pairs, consisting of $650$ evidence. Out of all the samples, the rater agrees with \texttt{GPT-4} 84.7\% of the time.\footnote{To further validate this number, we check agreement between two expert annotators. On $400$ examples, a $94\%$ agreement is reached. This indicates that the task is less subjective than general relevance tasks which tend to have a lower agreement, explaining the relatively high human-GPT agreement.} Specifically, they disagreed with the model on $141$ cases of ``relevant'' and only $57$ cases of ``not relevant''.

\subsection{Natural-language Query Generation} \label{sec:nl_query_generation} We generate natural sounding queries by providing \texttt{GPT-4} the ``list of'' page title and instructing the model to phrase a natural-language query. For details and examples see \cref{app:natural_lang_gen}.

\subsection{\ourdataset{} Overview} \label{sec:statistics} 
The final dataset comprises $1,196$ queries, encompassing $60,333$ evidence in total. There are $50.44$ evidence per query on average, and a median of $22$, ranging from a minimum of $5$ to a maximum of $682$ evidence. On average, each group member contributes about $2$ evidence to a query, with $61.8\%$ of the evidence coming from articles other than the members' own articles. The average number of members per query stands at $23.71$. We note that it is possible for some members to not contribute any evidence to a query, for example, when the evidence is not in the introduction. In \cref{tab:members_in_query_and_avg_evidence} we show the members and evidence distributions, and the relation between the number of members and number of evidence mapped to a query.

As accustomed with new datasets, we benchmark \ourdataset{} on the evidence retrieval task, where all evidence should be retrieved for a given query. Results are reported and discussed in \cref{app:dataset_baselines}.

\begin{table}[h]
\centering
\begin{tabular}{c|c|c}
\textbf{\# Members} & \textbf{Avg \# Evidence} & \textbf{\# Queries} \\ 
\hline

1-10      & 25.5 & 558  \\
11-20     & 32.0 & 282  \\
21-50     & 69.8 & 236  \\
51-100    &  109.7 & 77   \\
100+      &  281.2 & 43   \\

\end{tabular}
\caption{Dataset distribution average number of evidence over number of queries divided to buckets by number of set members.}
\label{tab:members_in_query_and_avg_evidence}
\end{table}

\section{Experimental Study} \label{sec:exp_study}
With our evaluation set ready, we can address the questions we put forth in the beginning. 
We experiment to examine the widespread practice of considering only a single evidence per query, and explore whether rankings stabilize as falsely labeled negatives decrease when adding more labeled evidence.

\subsection{Setup} \label{sec:experiment_setup}
\paragraph{Systems.} To ensure our analysis is unbiased towards a specific retrieval paradigm, we utilize the Pyserini information retrieval toolkit \cite{pyserini} to experiment across twelve diverse, out-of-the-box systems: five sparse, four dense, and three hybrid systems. (1) In the sparse category; BM25 \cite{bm25}, QLD~\cite{qld}, UniCoil~\cite{unicoil}, SPLADEv2 \cite{spladev2} and SPLADE++~\cite{spladepp}. (2) For the dense methods; DPR \cite{dpr}, coCondenser \cite{cocondenser}, RetroMAE-distill \cite{RetroMAE}, and TCT-Colbert-V2~\cite{tctcolbert}. (3) In the hybrid category; TCT-Colbert-V2-Hybrid~\cite{tctcolbert}, coCondenser-Hybrid, and RetroMAE-Hybrid. Further details regarding the systems can be found in \cref{app:further_details_experimental_study}.

\paragraph{Evaluation metrics.} Needing a metric to quantify the ability of systems to retrieve multiple evidence, we opt to use \textit{recall@$k$} as this is a simple, common metric for this task. For brevity, we report \textit{recall@20} in the main paper, and show results on \textit{recall@5}, \textit{recall@50}, and \textit{recall@100} in \cref{app:more_results}. We note that other \textit{k} values show similar trends to \textit{k=20}, and conclusions drawn in this paper generalize to other \textit{k} values reported as well. Other suitable metrics (NDCG, MAP, R-precision) are discussed and reported in \cref{app:dataset_baselines}. After evaluating the performance of each system, we are interested in comparing the recall-based ranking of systems to quantify the gap between the partially- and fully-annotated settings. We utilize Kendall-$\tau$ \citep{kendall_tau}, which can intuitively be understood as a measure of similarity between two ranking orders. This metric evaluates the number of pairwise agreements (concordant pairs) versus disagreements (discordant pairs) in the ranking order of systems between the two settings. A high Kendall-$\tau$ score (close to $1$) indicates a strong correlation, signifying that the rankings in the partially- and fully-annotated settings are similar, whereas a low score (close to $-1$) suggests major differences. Specifically, if we have \( n \) systems, and \( C \) is the number of concordant pairs while \( D \) is the number of discordant pairs, then Kendall-$\tau$ is given by the formula \( \tau = \frac{C - D}{{n \choose 2}} \), where ${n \choose 2}$ is the total number of possible pairs. 
In addition to the vanilla Kendall-$\tau$, we also report the probability of observing a discordant pair, denoted as the \emph{Error-rate}, as it is a more intuitive metric. Formally it is defined as:

$$\text{\textit{Error-rate}} = 100\cdot \frac{D}{{n \choose 2}} = 100 \cdot \frac{1 - \tau}{2}.$$

\subsection{Is the single-relevant setup reliable?}\label{sec:single_evidence}
To assess the single-relevant setup, we start by randomly sampling an evidence for each query. We evaluate each system on the formed single-relevant evaluation set and compare the resulting system ranking to the ground-truth ranking formed using the fully-annotated dataset. To mitigate the randomness, we run this experiment $1,000$ times, and find that the mean ($\pm$ std) Kendall-$\tau$ value is $0.936$ ($\pm 0.038$), translating to an error-rate of $3.2\%$. These numbers suggest that sampling a random evidence for each query leads to reliable results. Unfortunately, in order to properly randomly sample an evidence, one would need to annotate a non-feasible amount of passages in most datasets.\footnote{For example, in the 2020 TREC challenge \cite{trec2020}, operating on the MS-MARCO \cite{msmarco} dataset, $11,386$ relevant passages were found for $54$ queries, an average of $210$ per query. In \cref{app:trec_coverage_simulation} we estimate these are only $\sim50\%$ of the actual relevant passages leading to roughly $500$ per query. Given the corpus size, $\sim8M$ passages, one would need $\sim16K$ annotations on average to find a single relevant passage randomly for a \emph{single} query.}





\begin{table}[ht]
\centering
\begin{tabular}{c|c|c}
\textbf{Selection} & \textbf{$\tau$-similarity}& \textbf{Error-rate (\%)} \\ 
\hline

Random & \textbf{0.936} & \hspace{0.2cm} \textbf{3.20} \ \\ \hline
Most popular      & 0.696 & 15.10\ \\
Longest      & 0.545 & 22.75\ \\
Shortest      & 0.696 & 15.10\ \\ \hline
System-based & 0.616 & 19.20\ \\

\end{tabular}
\caption{Kendall-$\tau$ similarities and error-rate for the different biases in a single-annotation setup.}
\label{tab:kendall_tau_single_relevant_20}
\end{table}

\begin{figure}[h!]
    \includegraphics[width=7cm]{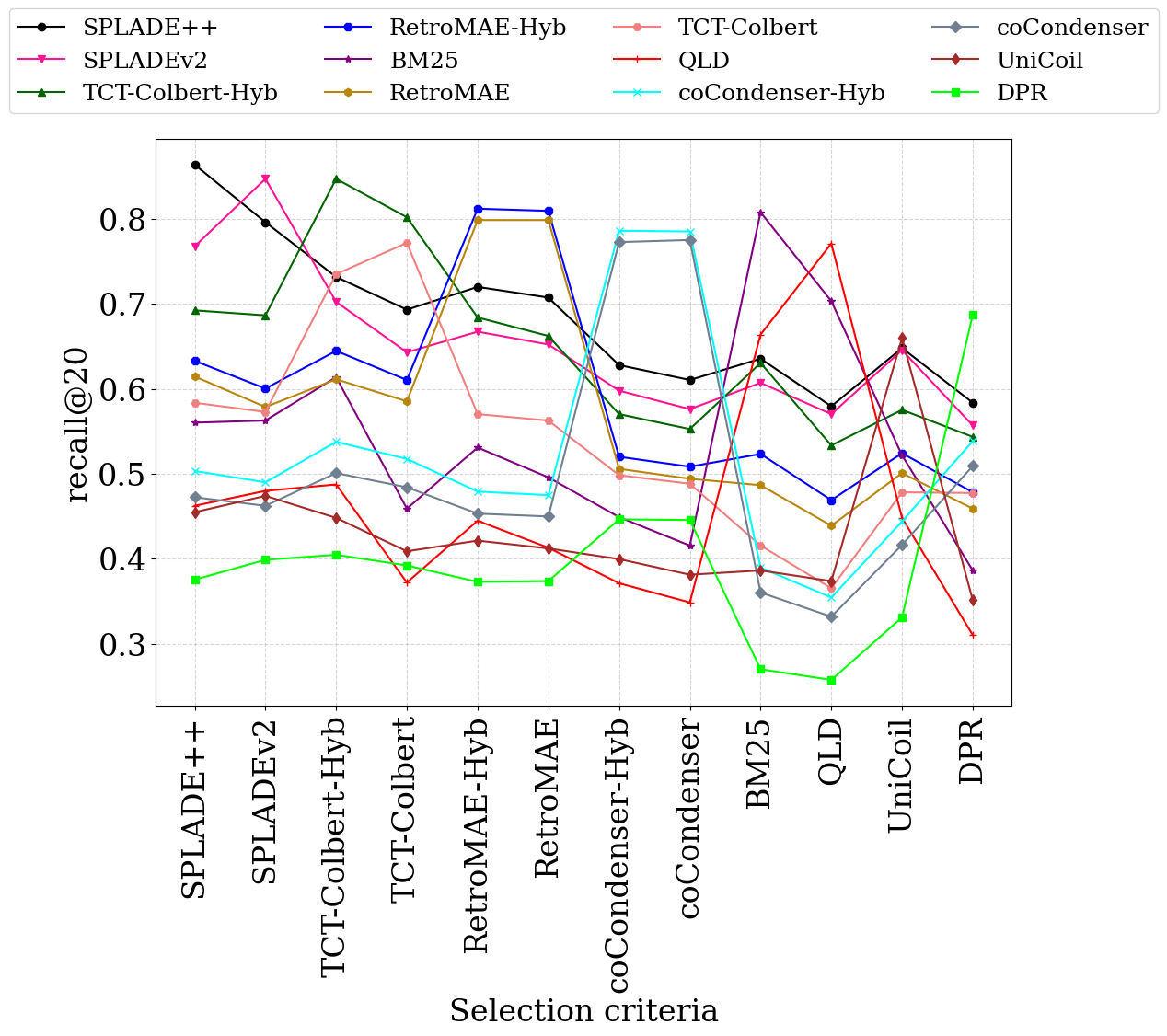}
      {\caption{Selection techniques for a single-relevant setting. The x-axis denotes systems used to select passages for annotation. Each tick represents the performance of systems on the same dataset with different annotations. An intersection demonstrates a swap in rankings.}
    \label{fig:swaps}}
\end{figure}

In practice, some method is used to select the passages sent for annotation. This method is usually biased\footnote{For example, it has been shown that models tend to suffer from popularity bias \citep{survivorship_bias_msmarco} and that sparse methods tend to prefer longer texts over shorter ones while a human annotator is likely to prefer shorter texts.}. To determine whether selecting an evidence in a biased manner is problematic or not, we explore $3$ biases: \emph{most popular} selects the most popular\footnote{We define popularity as the number of times an article is referenced, which can be derived using the ``What Links Here'' feature from \cref{sec:candidate_collection}.} evidence for each query. We also consider a length-selection approach, which considers the number of words in a given passage, by selecting the \emph{longest} and \emph{shortest} evidence available for each query. Results are presented in \cref{tab:kendall_tau_single_relevant_20}. It can be seen that as opposed to random selection, in the more likely scenario of a biased selection the error-rate is much higher, suggesting that the single-relevant setting is unreliable. A popular technique for sampling passages for annotation is using an existing retrieval system, and annotating passages in the order they are retrieved until a relevant passage is found. We simulate this by considering each of our $12$ considered retrievers as the base system. We then evaluate all of the systems on the $12$ formed evaluation sets. Results are plotted in \cref{fig:swaps}. The graph shows that the selection technique, used to pick which passages are annotated, has a major effect on the systems' measured performance \emph{and} on the ranking of the different systems. For example, when choosing evidence using BM-25, QLD is ranked as the best system (excluding BM-25 itself), while when choosing evidence using either coCondenser, coCondenser-Hybrid, DPR or TCT-Colbert, QLD is the worst performing system. For other systems selecting evidence, it is ranked somewhere in between. When comparing the $12$ rankings formed using these evaluation sets to the ranking formed by the completely annotated dataset, the average Kendall-$\tau$ score computed is $0.616$, translating to an average error-rate of $19.2\%$.\footnote{We eliminate the system used to select the evidence from the computation, as it generates artificial swaps. For example when computing the Kendall-$\tau$ for the ranking formed by choosing the first evidence as ranked by BM-25, Kendall-$\tau$ is computed on the ranking of all except BM-25.} \cref{tab:kendall_tau_single_relevant_20} indicates that system-based selection is indeed closer to biased selection than it is to random selection. In summary, the experiments presented in this section show that while random selection of evidence can lead to reliable results in the single-relevant scenario, the more realistic case (where the annotated evidence is not randomly selected) is prone to generating misleading results and ranking of systems.

\subsection{Is the single-relevant scenario enough when systems are significantly separated?} \label{sec:buckets_exp}
After establishing that there are cases where the single-relevant scenario is not reliable, we ask in what cases it can be sufficient. To explore this, we first define buckets of pairs of systems as follows. 
A pair of systems $(A,B)$ is in a $[p_{min}, p_{max})$ bucket if $A$ is better performing than $B$, and the statistical significance computation for the difference between these two systems leads to a p-value of at least $p_{min}$ and at most $p_{max}$, using a relative t-test, as computed on the fully annotated evaluation set.
We then repeat the final experiment described in \cref{sec:single_evidence}, but when calculating Kendall-$\tau$ and it's error-rate we only consider pairs of systems that fall in some bucket.
We denote this measure as partial-Kendall-$\tau$.\footnote{We opt to use Kendall-$\tau$ due to its simplicity, yet it does not accurately capture all the intricacies of ranking system performance. More details on this and an involved metric, taking into account the significance of differences between systems, is presented in Appendix~\ref{app:conc}. Results using this metric validate our choice of Kendall-$\tau$.}
We consider $3$ buckets: $[0,0.01)$ represents systems with very low p-values, meaning they are very far apart in performance, hence should be easier to order correctly. $[0.01,0.05)$ represents systems with a significant, yet not extreme difference. The final bucket, $[0.05,1)$, contains pairs of systems that do not differentiate in a statistically significant way.
Results are shown in \cref{tab:kendall_tau_single_relevant_buckets}. 
We observe that, as expected, the error-rate drops when a bucket represents a smaller p-value, indicating higher significance that the systems are ordered correctly.




\begin{table}[h]
\centering
\begin{tabular}{cc|c|c}
\textbf{$p_{min}$} & \textbf{$p_{max}$} & \textbf{partial-$\tau$}& \textbf{Error-rate (\%)} \\ 
\hline

0.0 & 0.01 & \textbf{0.658} & \textbf{17.1}  \\ 
0.01 & 0.05 & 0.333 & 33.3 \\
0.05 & 1.0 & 0.0 & 50.0

\end{tabular}
\caption{Partial-Kendall-$\tau$ similarity (defined in \cref{sec:buckets_exp}, denoted partial-$\tau$) and Error-rate computed on pairs of systems that belong to the [$p_{min}$, $p_{max}$) bucket.}
\label{tab:kendall_tau_single_relevant_buckets}
\end{table}

\subsection{Do rankings stabilize as falsely labeled negatives decrease?} \label{sec:effect_of_number_of_relevant_passages}
\begin{figure}
    \includegraphics[width=7cm]{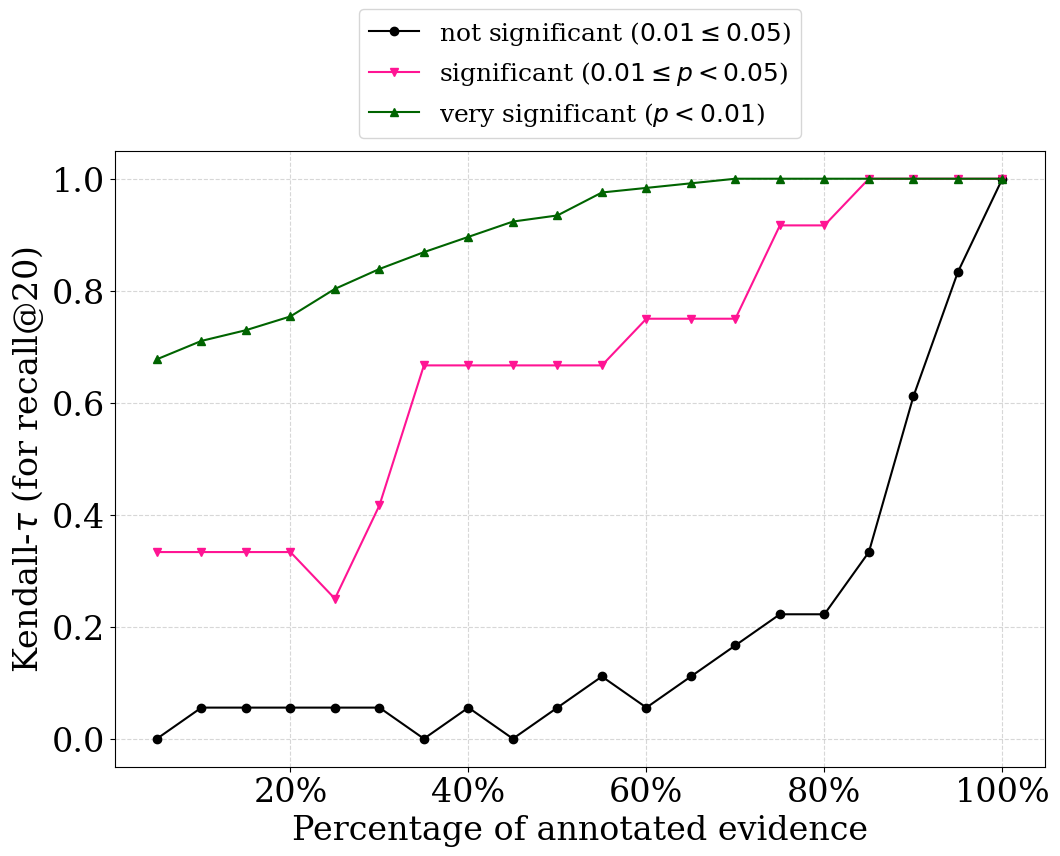}
    {\caption{Partial-Kendall-$\tau$ between rankings of systems with $k$ percent annotations and ranking with all evidence, using \recall{20}. System pairs are divided into 3 buckets as described in \cref{sec:buckets_exp}.}\label{fig:kendall_tau_recall_20}}
\end{figure}

Taking the evidence chosen using the different systems as discussed in \cref{sec:single_evidence}, we gradually add a fraction of annotated evidence for each query in the evaluation set. We then evaluate the systems on each partially annotated dataset by comparing the ranking achieved to the fully annotated evaluation set. We divide pairs of systems into buckets based on their p-values, as described in \cref{sec:buckets_exp}, and for each percentile we average results across the different system pairs falling within each bucket. Results are presented in \cref{fig:kendall_tau_recall_20}. Depending on the significance of the difference between systems, results show a different portion of evidence needs to be annotated in order to achieve the correct order.
For example, if we are aiming at a $\sim0.8$ Kendall-$\tau$ score, representing a $\sim10\%$ error-rate, for very significant pairs of systems acquiring $\sim20\%$ of the positives should suffice, while for systems with a non-significant difference between them, almost all positives are needed.

\section{Related Work} 
Our work builds on previous efforts in benchmark creations in multi-answer and multi-evidence settings and the complete annotation setting. Below, we detail how our work relates to both.

\paragraph{Multi-answer retrieval.} 
QAMParI \citep{qampari} introduce a benchmark of questions with multiple answers extracted from lists in Wikipedia, and Quest \citep{quest} is a dataset with queries containing implicit set operations based on Wikipedia category names. Both limit evidence collection to the Wikipedia article of the answer. In contrast, our goal is to identify all relevant evidence for each answer, including other Wikipedia articles. RomQA \citep{romqa} curates a large multi-evidence and multi-answer benchmark derived from the Wikidata knowledge graph with the goal of challenging the retriever and QA model. Although RomQA provides a large number of evidence, they do not aim for complete annotation nor to understand the negative effect of evaluation with partial annotations. Our paths diverge in that they seek to evaluate QA models and we aim to understand the effects of partial annotations on retriever evaluation, and to collect \textit{all} evidence for each answer.

\paragraph{Exhaustive annotation.}
TREC Deep Learning \citep{trec2019, trec2020, trec2021, trec2022, trec2023} is a yearly effort to completely-annotate queries for passage retrieval from the MS-Marco benchmark \citep{msmarco}. Since annotating the entirety of MS-MARCO is unrealistic (\textasciitilde 1M queries and \textasciitilde 8.8M passages), they conduct a competition where participants submit the results of their retrievers. Then, the results are pooled and their relevancy is evaluated.
However, manual evaluation is a non-scalable approach, and over a span of five years (2019--2023) only 312 queries were annotated. In addition, exhaustiveness is unlikely as previously observed in \cite{how_reliable_are_the_results_of_large_scale_info_retrieval_experiments} and further corroborated in Appendix~\ref{app:trec_coverage_simulation}.
NERetrieve \citep{neretrieve} shares our aspiration for a completely-annotated dataset. It proposes a retrieval-based NER task that creates a Wikipedia-based dataset where entity types function as queries and relevant passages contain a span that mentions instances of the entities (e.g., ``Dinosaurs'' is an entity type and ``Velociraptor'' is an instance of it). With some similarity to our process, they collect candidates by relaxed matching of mentions of entities in documents that reference them (on DBPedia's link-graph \citep{dbpedia}), and then use a classifier to filter out cases that do not match their query. However, our work annotates evidence and not simply mentions of entities in a passage. Moreover, in addition to creating an exhaustively annotated dataset, we study the effects of partial annotation.

\section{Conclusions}
In this work we question whether the lack of rigorous annotation in modern retrieval datasets results in false conclusions. To answer this, we create \ourdataset{} from Wikipedia. \ourdataset{} aspires to collect \textit{all} relevant passages in the corpus for each query, a property made possible due to Wikipedia's unique structure. We use \ourdataset{} to explore the impact of evaluating systems on datasets riddled with falsely labeled negatives; 
We demonstrate that evaluation based on queries with a single annotated relevant passage is highly dependent on the passages selected for annotation, unless one system is significantly superior to all others. 
We also show that the number of annotations required to stabilize the rankings is a factor of the difference in performance between systems.
We conclude that there is a clear efficiency-reliability curve when it comes to the amount of annotations invested in a retrieval evaluation set, and that when picking the correct spot on this curve considerations should include the estimated difference between the systems in question and the method used to choose the passages sent to annotation.
We show that the commonly used TREC-style evaluation method fails to find a significant portion of the relevant passages in \ourdataset{}, suggesting that using this annotation approach on \ourdataset{} would lead to a non-negligible error rate. If it's possible, our recommendation for other datasets would be to estimate the coverage of the TREC method before using it for evaluation. Otherwise, its results should be taken with a grain-of-salt.
Finally, our dataset opens a new avenue for research, both as a test-bed for evaluation studies, as well as evaluation in a high-recall setting.

\section*{Limitations}

\textbf{Generalization of conclusions.} We (and many before us) believe that in order to properly evaluate retrieval systems, the community should \emph{strive} to collect all (or most) relevant passages. We believe this is true for many different datasets and scenarios. Having said that, showing this explicitly requires to completely annotate datasets, which is hard and expensive. Furthermore, our dataset collection method does not generalize to other corpora as it highly relies on the Wikipedia structure (specifically, on the "list of" pages). Therefore, while we do believe that most of our conclusions can generalize to many other datasets, technically we could show them only on the dataset we used.

\textbf{Exhaustiveness.} Our evidence identification process is automated by \texttt{GPT-4}, the current state-of-the-art for text analysis. 
Despite achieving high agreement with human annotators, it 
is not perfect. 
Furthermore, even with a flawless model, computing the relevance of \emph{all} passages in Wikipedia for each member in each query would have resulted in millions of inferences, which would have made the creation of this dataset unfathomably expensive. We thus make the (sensible) assumption that a passage with evidence must contain a link to the article of the entity. 
It is possible some evidence were never collected, as analyzed in \cref{sec:eval_of_dataset_construction}.

\textbf{Data evaluation compatibility.} Our dataset is made of set-queries with multiple members (translating to multiple answers in the QA setting). In such cases, systems are usually evaluated using datasets containing a single relevant \textit{per answer}. In \cref{sec:single_evidence} we evaluate and draw conclusions using a single positive \textit{per query}. We do so in order to draw conclusions regarding cases where single positives per query are used, but in practice these datasets usually contain \textit{single-answer} queries (e.g. MS-MARCO). While we do believe our conclusions generalize to this case, it would have been more accurate to use such a single-answer-per-query dataset. Unfortunately, collecting such a fully annotated dataset is not trivial.

\section*{Ethics Statement}
\paragraph{Automatic annotation.} Since our annotation is automatic, it is model-dependent. This means it is vulnerable to the model's biases. As a result, it may fail to attribute evidence to a query if a candidate is under-represented in the model's training data. This might cause \ourdataset{} to miss out on evidence that belongs to some under-represented group.

\paragraph{Rater details.} To collect annotations on our dataset, we used Amazon Mechanical Turk (AMT). All raters had the following qualifications: (1) over
5,000 completed HITs; (2) 99\% approval rate or
higher; (3) Native English speakers from England,
New Zealand, Canada, Australia, or United States.
Raters were paid \$0.07 per HIT, and on average, \$20 an hour. In addition, raters that performed the task well were given bonuses that reached double pay.

\paragraph{Annotation collection and usage policy.} Raters were notified that their annotations are intended for research use in the field of Natural Language Processing and Information Retrieval, and will ultimately be shared publicly. The task and collected annotations were objective and excluded personal information. Moreover, all data sources for the study were publicly accessible.

\paragraph{Computing resources.} We used only modest computing resources. For both, the dataset creation and the experimentation, we used a single Amazon-EC2-g5.4xlarge instance for 200 hours, which costs \$1.6 per hour. For the annotation of the passages, and creation of the natural-language queries, we utilized \texttt{GPT-4-1106-preview}, which at the time of writing, is priced at \$0.01 for 1K input tokens, and \$0.03 for 1K output tokens. In total, we paid \textasciitilde \$3,000 for our use of the model.

\section*{Acknowledgements}
This project received funding from the European Research Council (ERC) under the European Union's Horizon 2020 research and innovation programme, grant agreement No. 802774 (iEXTRACT).

\clearpage
\bibliography{anthology,main}

\begin{thebibliography}{48}
\expandafter\ifx\csname natexlab\endcsname\relax\def\natexlab#1{#1}\fi

\bibitem[{Amouyal et~al.(2023)Amouyal, Wolfson, Rubin, Yoran, Herzig, and Berant}]{qampari}
Samuel Amouyal, Tomer Wolfson, Ohad Rubin, Ori Yoran, Jonathan Herzig, and Jonathan Berant. 2023.
\newblock \href {https://aclanthology.org/2023.gem-1.9} {{QAMPARI}: A benchmark for open-domain questions with many answers}.
\newblock In \emph{Proceedings of the Third Workshop on Natural Language Generation, Evaluation, and Metrics (GEM)}, pages 97--110, Singapore. Association for Computational Linguistics.

\bibitem[{Bajaj et~al.(2018)Bajaj, Campos, Craswell, Deng, Gao, Liu, Majumder, McNamara, Mitra, Nguyen, Rosenberg, Song, Stoica, Tiwary, and Wang}]{msmarco}
Payal Bajaj, Daniel Campos, Nick Craswell, Li~Deng, Jianfeng Gao, Xiaodong Liu, Rangan Majumder, Andrew McNamara, Bhaskar Mitra, Tri Nguyen, Mir Rosenberg, Xia Song, Alina Stoica, Saurabh Tiwary, and Tong Wang. 2018.
\newblock \href {http://arxiv.org/abs/1611.09268} {Ms marco: A human generated machine reading comprehension dataset}.

\bibitem[{Bekoulis et~al.(2021)Bekoulis, Papagiannopoulou, and Deligiannis}]{bekoulis2021review}
Giannis Bekoulis, Christina Papagiannopoulou, and Nikos Deligiannis. 2021.
\newblock \href {https://doi.org/10.1145/3485127} {A review on fact extraction and verification}.
\newblock \emph{ACM Comput. Surv.}, 55(1).

\bibitem[{Buckley et~al.(2007)Buckley, Dimmick, Soboroff, and Voorhees}]{bias_of_pooling}
C~Buckley, Darrin Dimmick, Ian Soboroff, and Ellen Voorhees. 2007.
\newblock \href {https://tsapps.nist.gov/publication/get_pdf.cfm?pub_id=51236} {Bias and the limits of pooling for large collections}.

\bibitem[{Buckley and Voorhees(2004)}]{retrieval_with_incomplete_info}
Chris Buckley and Ellen~M. Voorhees. 2004.
\newblock \href {https://doi.org/10.1145/1008992.1009000} {Retrieval evaluation with incomplete information}.
\newblock In \emph{Proceedings of the 27th Annual International ACM SIGIR Conference on Research and Development in Information Retrieval}, SIGIR '04, page 25–32, New York, NY, USA. Association for Computing Machinery.

\bibitem[{Cai et~al.(2022)Cai, Wang, Liu, and Shi}]{cai2022recent}
Deng Cai, Yan Wang, Lemao Liu, and Shuming Shi. 2022.
\newblock \href {https://doi.org/10.1145/3477495.3532682} {Recent advances in retrieval-augmented text generation}.
\newblock In \emph{Proceedings of the 45th International ACM SIGIR Conference on Research and Development in Information Retrieval}, SIGIR '22, page 3417–3419, New York, NY, USA. Association for Computing Machinery.

\bibitem[{Callan(1994)}]{callan1994passage}
James~P. Callan. 1994.
\newblock Passage-level evidence in document retrieval.
\newblock In \emph{Proceedings of the 17th Annual International ACM SIGIR Conference on Research and Development in Information Retrieval}, SIGIR '94, page 302–310, Berlin, Heidelberg. Springer-Verlag.

\bibitem[{Craswell et~al.(2021)Craswell, Mitra, Yilmaz, and Campos}]{trec2020}
Nick Craswell, Bhaskar Mitra, Emine Yilmaz, and Daniel Campos. 2021.
\newblock \href {https://www.microsoft.com/en-us/research/publication/overview-of-the-trec-2020-deep-learning-track/} {Overview of the trec 2020 deep learning track}.
\newblock In \emph{Text REtrieval Conference (TREC)}. TREC.

\bibitem[{Craswell et~al.(2022)Craswell, Mitra, Yilmaz, Campos, and Lin}]{trec2021}
Nick Craswell, Bhaskar Mitra, Emine Yilmaz, Daniel Campos, and Jimmy Lin. 2022.
\newblock \href {https://www.microsoft.com/en-us/research/publication/overview-of-the-trec-2021-deep-learning-track/} {Overview of the trec 2021 deep learning track}.
\newblock In \emph{Text REtrieval Conference (TREC)}. NIST, TREC.

\bibitem[{Craswell et~al.(2023)Craswell, Mitra, Yilmaz, Campos, Lin, Voorhees, and Soboroff}]{trec2022}
Nick Craswell, Bhaskar Mitra, Emine Yilmaz, Daniel Campos, Jimmy Lin, Ellen~M. Voorhees, and Ian Soboroff. 2023.
\newblock \href {https://www.microsoft.com/en-us/research/publication/overview-of-the-trec-2022-deep-learning-track/} {Overview of the trec 2022 deep learning track}.
\newblock In \emph{Text REtrieval Conference (TREC)}. NIST, TREC.

\bibitem[{Craswell et~al.(2020)Craswell, Mitra, Yilmaz, Campos, and Voorhees}]{trec2019}
Nick Craswell, Bhaskar Mitra, Emine Yilmaz, Daniel Campos, and Ellen~M. Voorhees. 2020.
\newblock \href {http://arxiv.org/abs/2003.07820} {Overview of the trec 2019 deep learning track}.

\bibitem[{Craswell et~al.(2024)Craswell, Mitra, Yilmaz, Rahmani, Campos, Lin, Voorhees, and Soboroff}]{trec2023}
Nick Craswell, Bhaskar Mitra, Emine Yilmaz, Hossein~A. Rahmani, Daniel Campos, Jimmy Lin, Ellen~M. Voorhees, and Ian Soboroff. 2024.
\newblock \href {https://www.microsoft.com/en-us/research/publication/overview-of-the-trec-2023-deep-learning-track/} {Overview of the trec 2023 deep learning track}.
\newblock In \emph{Text REtrieval Conference (TREC)}. NIST, TREC.

\bibitem[{Douze et~al.(2024)Douze, Guzhva, Deng, Johnson, Szilvasy, Mazaré, Lomeli, Hosseini, and Jégou}]{faiss}
Matthijs Douze, Alexandr Guzhva, Chengqi Deng, Jeff Johnson, Gergely Szilvasy, Pierre-Emmanuel Mazaré, Maria Lomeli, Lucas Hosseini, and Hervé Jégou. 2024.
\newblock \href {http://arxiv.org/abs/2401.08281} {The faiss library}.

\bibitem[{Formal et~al.(2022)Formal, Lassance, Piwowarski, and Clinchant}]{spladepp}
Thibault Formal, Carlos Lassance, Benjamin Piwowarski, and St\'{e}phane Clinchant. 2022.
\newblock \href {https://doi.org/10.1145/3477495.3531857} {From distillation to hard negative sampling: Making sparse neural ir models more effective}.
\newblock In \emph{Proceedings of the 45th International ACM SIGIR Conference on Research and Development in Information Retrieval}, SIGIR '22, page 2353–2359, New York, NY, USA. Association for Computing Machinery.

\bibitem[{Formal et~al.(2021)Formal, Lassance, Piwowarski, and Clinchant}]{spladev2}
Thibault Formal, Carlos Lassance, Benjamin Piwowarski, and Stéphane Clinchant. 2021.
\newblock \href {http://arxiv.org/abs/2109.10086} {Splade v2: Sparse lexical and expansion model for information retrieval}.

\bibitem[{Gao and Callan(2022)}]{cocondenser}
Luyu Gao and Jamie Callan. 2022.
\newblock \href {https://doi.org/10.18653/v1/2022.acl-long.203} {Unsupervised corpus aware language model pre-training for dense passage retrieval}.
\newblock In \emph{Proceedings of the 60th Annual Meeting of the Association for Computational Linguistics (Volume 1: Long Papers)}, pages 2843--2853, Dublin, Ireland. Association for Computational Linguistics.

\bibitem[{Gupta and MacAvaney(2022)}]{survivorship_bias_msmarco}
Prashansa Gupta and Sean MacAvaney. 2022.
\newblock \href {https://doi.org/10.1145/3477495.3531832} {On survivorship bias in ms marco}.
\newblock In \emph{Proceedings of the 45th International ACM SIGIR Conference on Research and Development in Information Retrieval}, SIGIR ’22. ACM.

\bibitem[{Karpukhin et~al.(2020)Karpukhin, Oguz, Min, Lewis, Wu, Edunov, Chen, and Yih}]{dpr}
Vladimir Karpukhin, Barlas Oguz, Sewon Min, Patrick Lewis, Ledell Wu, Sergey Edunov, Danqi Chen, and Wen-tau Yih. 2020.
\newblock \href {https://doi.org/10.18653/v1/2020.emnlp-main.550} {Dense passage retrieval for open-domain question answering}.
\newblock In \emph{Proceedings of the 2020 Conference on Empirical Methods in Natural Language Processing (EMNLP)}, pages 6769--6781, Online. Association for Computational Linguistics.

\bibitem[{Kaszkiel and Zobel(1997)}]{kaszkiel1997passage}
Marcin Kaszkiel and Justin Zobel. 1997.
\newblock Passage retrieval revisited.
\newblock In \emph{ACM SIGIR Forum}, volume~31, pages 178--185. ACM New York, NY, USA.

\bibitem[{Katz et~al.(2023)Katz, Vetzler, Cohen, and Goldberg}]{neretrieve}
Uri Katz, Matan Vetzler, Amir Cohen, and Yoav Goldberg. 2023.
\newblock \href {https://doi.org/10.18653/v1/2023.findings-emnlp.218} {{NER}etrieve: Dataset for next generation named entity recognition and retrieval}.
\newblock In \emph{Findings of the Association for Computational Linguistics: EMNLP 2023}, pages 3340--3354, Singapore. Association for Computational Linguistics.

\bibitem[{Keenan et~al.(2001)Keenan, Smeaton, and Keogh}]{effect_of_pool_depth_trec_eval}
Sabrina Keenan, Alan~F. Smeaton, and Gary Keogh. 2001.
\newblock \href {https://doi.org/10.1002/asi.1096.abs} {The effect of pool depth on system evaluation in trec}.
\newblock \emph{J. Am. Soc. Inf. Sci. Technol.}, 52(7):570–574.

\bibitem[{Kendall(1938)}]{kendall_tau}
M.~G. Kendall. 1938.
\newblock \href {http://www.jstor.org/stable/2332226} {A new measure of rank correlation}.
\newblock \emph{Biometrika}, 30(1/2):81--93.

\bibitem[{Kendall(1945)}]{kendall1945treatment}
Maurice~G Kendall. 1945.
\newblock The treatment of ties in ranking problems.
\newblock \emph{Biometrika}, 33(3):239--251.

\bibitem[{Kwiatkowski et~al.(2019)Kwiatkowski, Palomaki, Redfield, Collins, Parikh, Alberti, Epstein, Polosukhin, Devlin, Lee, Toutanova, Jones, Kelcey, Chang, Dai, Uszkoreit, Le, and Petrov}]{natural_questions}
Tom Kwiatkowski, Jennimaria Palomaki, Olivia Redfield, Michael Collins, Ankur Parikh, Chris Alberti, Danielle Epstein, Illia Polosukhin, Jacob Devlin, Kenton Lee, Kristina Toutanova, Llion Jones, Matthew Kelcey, Ming-Wei Chang, Andrew~M. Dai, Jakob Uszkoreit, Quoc Le, and Slav Petrov. 2019.
\newblock \href {https://doi.org/10.1162/tacl_a_00276} {Natural questions: A benchmark for question answering research}.
\newblock \emph{Transactions of the Association for Computational Linguistics}, 7:452--466.

\bibitem[{Lehmann et~al.(2015)Lehmann, Isele, Jakob, Jentzsch, Kontokostas, Mendes, Hellmann, Morsey, van Kleef, Auer, and Bizer}]{dbpedia}
Jens Lehmann, Robert Isele, Max Jakob, Anja Jentzsch, Dimitris Kontokostas, Pablo~N. Mendes, Sebastian Hellmann, Mohamed Morsey, Patrick van Kleef, S.~Auer, and Christian Bizer. 2015.
\newblock \href {https://api.semanticscholar.org/CorpusID:1181640} {Dbpedia - a large-scale, multilingual knowledge base extracted from wikipedia}.
\newblock \emph{Semantic Web}, 6:167--195.

\bibitem[{Lewis et~al.(2021)Lewis, Perez, Piktus, Petroni, Karpukhin, Goyal, Küttler, Lewis, tau Yih, Rocktäschel, Riedel, and Kiela}]{rag_paper}
Patrick Lewis, Ethan Perez, Aleksandra Piktus, Fabio Petroni, Vladimir Karpukhin, Naman Goyal, Heinrich Küttler, Mike Lewis, Wen tau Yih, Tim Rocktäschel, Sebastian Riedel, and Douwe Kiela. 2021.
\newblock \href {http://arxiv.org/abs/2005.11401} {Retrieval-augmented generation for knowledge-intensive nlp tasks}.

\bibitem[{Li et~al.(2022)Li, Su, Cai, Wang, and Liu}]{li2022survey}
Huayang Li, Yixuan Su, Deng Cai, Yan Wang, and Lemao Liu. 2022.
\newblock A survey on retrieval-augmented text generation.
\newblock \emph{arXiv preprint arXiv:2202.01110}.

\bibitem[{Lin and Ma(2021)}]{unicoil}
Jimmy Lin and Xueguang Ma. 2021.
\newblock A few brief notes on deepimpact, coil, and a conceptual framework for information retrieval techniques.
\newblock \emph{arXiv preprint arXiv:2106.14807}.

\bibitem[{Lin et~al.(2021{\natexlab{a}})Lin, Ma, Lin, Yang, Pradeep, and Nogueira}]{pyserini}
Jimmy Lin, Xueguang Ma, Sheng-Chieh Lin, Jheng-Hong Yang, Ronak Pradeep, and Rodrigo Nogueira. 2021{\natexlab{a}}.
\newblock \href {https://doi.org/10.1145/3404835.3463238} {Pyserini: A python toolkit for reproducible information retrieval research with sparse and dense representations}.
\newblock In \emph{Proceedings of the 44th International ACM SIGIR Conference on Research and Development in Information Retrieval}, SIGIR '21, page 2356–2362, New York, NY, USA. Association for Computing Machinery.

\bibitem[{Lin et~al.(2021{\natexlab{b}})Lin, Yang, and Lin}]{tctcolbert}
Sheng-Chieh Lin, Jheng-Hong Yang, and Jimmy Lin. 2021{\natexlab{b}}.
\newblock \href {https://doi.org/10.18653/v1/2021.repl4nlp-1.17} {In-batch negatives for knowledge distillation with tightly-coupled teachers for dense retrieval}.
\newblock In \emph{Proceedings of the 6th Workshop on Representation Learning for NLP (RepL4NLP-2021)}, pages 163--173, Online. Association for Computational Linguistics.

\bibitem[{Lu et~al.(2016)Lu, Moffat, and Culpepper}]{effect_of_pooling_on_ir_metrics}
Xiaolu Lu, Alistair Moffat, and J.~Shane Culpepper. 2016.
\newblock \href {https://doi.org/10.1007/s10791-016-9282-6} {The effect of pooling and evaluation depth on ir metrics}.
\newblock \emph{Inf. Retr.}, 19(4):416–445.

\bibitem[{MacAvaney and Soldaini(2023)}]{one_shot_labeling_relevance}
Sean MacAvaney and Luca Soldaini. 2023.
\newblock \href {https://doi.org/10.1145/3539618.3592032} {One-shot labeling for automatic relevance estimation}.
\newblock In \emph{Proceedings of the 46th International ACM SIGIR Conference on Research and Development in Information Retrieval}, SIGIR ’23. ACM.

\bibitem[{Malaviya et~al.(2023)Malaviya, Shaw, Chang, Lee, and Toutanova}]{quest}
Chaitanya Malaviya, Peter Shaw, Ming-Wei Chang, Kenton Lee, and Kristina Toutanova. 2023.
\newblock \href {http://arxiv.org/abs/2305.11694} {Quest: A retrieval dataset of entity-seeking queries with implicit set operations}.

\bibitem[{Mavi et~al.(2022)Mavi, Jangra, and Jatowt}]{mavi2022survey}
Vaibhav Mavi, Anubhav Jangra, and Adam Jatowt. 2022.
\newblock A survey on multi-hop question answering and generation.
\newblock \emph{arXiv preprint arXiv:2204.09140}.

\bibitem[{Murayama(2021)}]{murayama2021dataset}
Taichi Murayama. 2021.
\newblock Dataset of fake news detection and fact verification: a survey.
\newblock \emph{arXiv preprint arXiv:2111.03299}.

\bibitem[{Ram et~al.(2023)Ram, Levine, Dalmedigos, Muhlgay, Shashua, Leyton-Brown, and Shoham}]{ram2023context}
Ori Ram, Yoav Levine, Itay Dalmedigos, Dor Muhlgay, Amnon Shashua, Kevin Leyton-Brown, and Yoav Shoham. 2023.
\newblock \href {https://doi.org/10.1162/tacl_a_00605} {In-context retrieval-augmented language models}.
\newblock \emph{Transactions of the Association for Computational Linguistics}, 11:1316--1331.

\bibitem[{Robertson and Walker(1994)}]{bm25}
S.~E. Robertson and S.~Walker. 1994.
\newblock Some simple effective approximations to the 2-poisson model for probabilistic weighted retrieval.
\newblock In \emph{Proceedings of the 17th Annual International ACM SIGIR Conference on Research and Development in Information Retrieval}, SIGIR '94, page 232–241, Berlin, Heidelberg. Springer-Verlag.

\bibitem[{Rogers et~al.(2023)Rogers, Gardner, and Augenstein}]{rogers2023qa}
Anna Rogers, Matt Gardner, and Isabelle Augenstein. 2023.
\newblock \href {https://doi.org/10.1145/3560260} {Qa dataset explosion: A taxonomy of nlp resources for question answering and reading comprehension}.
\newblock \emph{ACM Comput. Surv.}, 55(10).

\bibitem[{Stuart(1953)}]{stuart1953estimation}
Alan Stuart. 1953.
\newblock The estimation and comparison of strengths of association in contingency tables.
\newblock \emph{Biometrika}, 40(1/2):105--110.

\bibitem[{Vallayil et~al.(2023)Vallayil, Nand, Yan, and Allende-Cid}]{vallayil2023explainability}
Manju Vallayil, Parma Nand, Wei~Qi Yan, and H{\'e}ctor Allende-Cid. 2023.
\newblock Explainability of automated fact verification systems: A comprehensive review.
\newblock \emph{Applied Sciences}, 13(23):12608.

\bibitem[{Vrandečić and Krötzsch(2014)}]{wikidata}
Denny Vrandečić and Markus Krötzsch. 2014.
\newblock \href {http://cacm.acm.org/magazines/2014/10/178785-wikidata/fulltext} {Wikidata: A free collaborative knowledge base}.
\newblock \emph{Communications of the ACM}, 57:78--85.

\bibitem[{Xiao et~al.(2022)Xiao, Liu, Shao, and Cao}]{RetroMAE}
Shitao Xiao, Zheng Liu, Yingxia Shao, and Zhao Cao. 2022.
\newblock \href {https://doi.org/10.18653/v1/2022.emnlp-main.35} {{R}etro{MAE}: Pre-training retrieval-oriented language models via masked auto-encoder}.
\newblock In \emph{Proceedings of the 2022 Conference on Empirical Methods in Natural Language Processing}, pages 538--548, Abu Dhabi, United Arab Emirates. Association for Computational Linguistics.

\bibitem[{Yilmaz and Aslam(2006)}]{infAP}
Emine Yilmaz and Javed~A. Aslam. 2006.
\newblock \href {https://doi.org/10.1145/1183614.1183633} {Estimating average precision with incomplete and imperfect judgments}.
\newblock In \emph{Proceedings of the 15th ACM International Conference on Information and Knowledge Management}, CIKM '06, page 102–111, New York, NY, USA. Association for Computing Machinery.

\bibitem[{Zhai and Lafferty(2001)}]{qld}
Chengxiang Zhai and John Lafferty. 2001.
\newblock \href {https://doi.org/10.1145/383952.384019} {A study of smoothing methods for language models applied to ad hoc information retrieval}.
\newblock In \emph{Proceedings of the 24th Annual International ACM SIGIR Conference on Research and Development in Information Retrieval}, SIGIR '01, page 334–342, New York, NY, USA. Association for Computing Machinery.

\bibitem[{Zhong et~al.(2022)Zhong, Shi, tau Yih, and Zettlemoyer}]{romqa}
Victor Zhong, Weijia Shi, Wen tau Yih, and Luke Zettlemoyer. 2022.
\newblock \href {http://arxiv.org/abs/2210.14353} {Romqa: A benchmark for robust, multi-evidence, multi-answer question answering}.

\bibitem[{Zhu et~al.(2021)Zhu, Lei, Wang, Zheng, Poria, and Chua}]{zhu2021retrieving}
Fengbin Zhu, Wenqiang Lei, Chao Wang, Jianming Zheng, Soujanya Poria, and Tat-Seng Chua. 2021.
\newblock Retrieving and reading: A comprehensive survey on open-domain question answering.
\newblock \emph{arXiv preprint arXiv:2101.00774}.

\bibitem[{Zobel(1998)}]{how_reliable_are_the_results_of_large_scale_info_retrieval_experiments}
Justin Zobel. 1998.
\newblock \href {https://api.semanticscholar.org/CorpusID:14804938} {How reliable are the results of large-scale information retrieval experiments?}
\newblock In \emph{Annual International ACM SIGIR Conference on Research and Development in Information Retrieval}.

\bibitem[{Zobel et~al.(1995)Zobel, Moffat, Wilkinson, and Sacks-Davis}]{zobel1995efficient}
Justin Zobel, Alistair Moffat, Ross Wilkinson, and Ron Sacks-Davis. 1995.
\newblock \href {https://doi.org/https://doi.org/10.1016/0306-4573(94)00052-5} {Efficient retrieval of partial documents}.
\newblock \emph{Information Processing \& Management}, 31(3):361--377.
\newblock The Second Text Retrieval Conference (TREC-2).

\end{thebibliography}

\clearpage
\appendix


\begin{table*}[t]
\centering
\scalebox{0.85}{
\small
\begin{tabular}{l|c|c|c|c|c|c|c|c|c|c|c|c|c}
\multirow{2}{*}{\textbf{System}} & \multicolumn{4}{c|}{\textbf{Recall@k}} & \multicolumn{4}{c|}{\textbf{NDCG@k}}  & \multicolumn{4}{c|}{\textbf{MAP@k}} & \multirow{2}{*}{\textbf{R-precision}} \\ 
\cline{2-13}
 & \textbf{5} & \textbf{20} & \textbf{50} & \textbf{100} & \textbf{5} & \textbf{20} & \textbf{50} & \textbf{100} & \textbf{5} & \textbf{20} & \textbf{50} & \textbf{100} &  \\ 
\hline
SPLADE++ & \bf{9.43} & \bf{24.11} & \bf{36.02} & \bf{45.16} & \bf{38.17} & \bf{36.54} & \bf{38.05} & \bf{40.56} & \bf{7.11} & \bf{15.0} & \bf{19.35} & \bf{21.72} & \bf{28.16} \\
SPLADEv2 & 7.82 & 21.21 & 33.29 & 43.34 & 32.09 & 31.43 & 33.78 & 37.00 & 5.74 & 12.20 & 16.03 & 18.27 & 24.82 \\
TCT-Colbert-Hybrid & 7.85 & 19.62 & 29.71 & 37.97 & 34.86 & 31.60 & 32.23 & 34.33 & 5.80 & 11.48 & 14.78 & 16.56 & 22.75 \\
bm25 & 6.65 & 17.46 & 27.54 & 35.76 & 28.93 & 27.20 & 28.62 & 31.13 & 4.76 & 9.76 & 12.83 & 14.61 & 20.86 \\
RetroMAE-Hybrid & 7.30 & 17.48 & 25.95 & 32.85 & 33.95 & 29.21 & 29.19 & 30.82 & 5.71 & 10.63 & 13.14 & 14.48 & 20.12 \\
RetroMAE & 7.03 & 16.62 & 24.78 & 31.61 & 32.71 & 27.98 & 27.94 & 29.61 & 5.47 & 10.05 & 12.38 & 13.66 & 19.29 \\
TCT-Colbert & 6.27 & 15.44 & 23.59 & 30.95 & 29.31 & 25.73 & 26.08 & 27.95 & 4.58 & 8.64 & 11.02 & 12.39 & 18.02 \\
CoCondenser-Hybrid & 5.28 & 14.81 & 24.25 & 32.88 & 22.13 & 21.87 & 23.96 & 26.89 & 3.41 & 6.82 & 9.10 & 10.63 & 16.78 \\
QLD & 5.49 & 13.96 & 23.56 & 31.96 & 24.54 & 21.71 & 23.63 & 26.55 & 3.77 & 7.07 & 9.51 & 11.13 & 16.56 \\
CoCondenser & 4.87 & 13.75 & 23.02 & 31.52 & 20.71 & 20.42 & 22.64 & 25.54 & 3.14 & 6.20 & 8.35 & 9.77 & 15.69 \\
Unicoil & 4.47 & 10.95 & 17.27 & 23.28 & 20.86 & 17.96 & 18.70 & 20.49 & 3.25 & 6.05 & 7.72 & 8.83 & 13.19 \\
DPR & 3.90 & 9.62 & 15.99 & 21.72 & 18.51 & 15.90 & 16.64 & 18.41 & 2.63 & 4.48 & 5.67 & 6.37 & 10.89 \\
\bottomrule
\end{tabular}}
\caption{Performance of a variety of baselines on \ourdataset{}. Recall, NDCG, and MAP are evaluated over four $k$ values: 5, 20, 50, and 100. The $k$ value in R-precision is the total number of evidence of a query, which changes from query to query.}
\label{app:dataset_baselines_tab}
\end{table*}

\label{app:appendix}
\section{Benchmarking \ourdataset{}} \label{app:dataset_baselines}
While tangential to this paper, the \ourdataset{} dataset allows us to benchmark the ability of existing retrieval models to perform on the full-recall retrieval setup, as it's coverage is very high as reported in \cref{sec:eval_of_dataset_construction}. This section describes this benchmark process.

\paragraph{Benchmark metrics.}
We select Recall, Normalized Discounted Cumulative Gain (NDCG) and Mean Average Precision (MAP). In addition, given that we possess complete evidence for every query, we can calculate R-precision-- a form of recall where $k$ varies for each query, determined by the specific total evidence count to that query. For instance, if a query corresponds to 40 pieces of evidence, then $k$ is set at 40. Achieving a perfect score means that the top 40 results are all evidence associated with the query.

\paragraph{Results.} Performance of all systems is shown in \cref{app:dataset_baselines_tab}, with SPLADE++ and SPLADEv2 performing best across all metrics. The scores suggest there is substantial room for improvement on our evidence retrieval task.  For example, the \recall{100} score indicates no system successfully retrieves even half of the evidence on average.

\section{Further Details: Experimental Study} \label{app:further_details_experimental_study}
To allow reproduction of our results, we detail the hyper-parameters used in our work. We utilize the Pyserini information retrieval toolkit \citep{pyserini} with the following settings for each system: \textbf{BM25} is employed using the standard Lucene index for indexing and retrieving results. Similarly, \textbf{QLD} is used but with the QLD reweighing option to refine the process. \textbf{UniCoil} embeddings are generated with the \emph{castorini/unicoil-noexp-msmarco-passage} encoder, and retrieval is conducted using Lucene search with the `impact' option to incorporate unicoil weights. \textbf{SPLADEv2} and \textbf{SPLADE++} follow a similar approach, where passages and queries are embedded using their respective official code repositories, and retrieval is performed using Lucene with the `impact' option. \textbf{DPR} involves embedding passages and queries with the \emph{facebook/dpr-ctx\_encoder-multiset-base} and \emph{facebook/dpr-question\_encoder-multiset-base} encoders, respectively, with retrieval via FAISS \citep{faiss}. \textbf{RetroMAE-distill} adopts a similar strategy, utilizing the \emph{Shitao/RetroMAE\_MSMARCO\_distill} encoder for both queries and passages. \textbf{TCT-Colbert-V2} also mirrors this approach but uses the \emph{castorini/tct\_colbert-v2-msmarco} encoder. \textbf{coCondenser} involves training document and query encoders on the Natural Questions dataset \citep{natural_questions} using the CoCondenser official code repository. Hybrid models such as \textbf{TCT-Colbert-V2-Hybrid}, \textbf{coCondenser-Hybrid}, and \textbf{RetroMAE-Hybrid} combine the strengths of BM25 with \textbf{TCT-Colbert-V2}, \textbf{coCondenser}, and \textbf{RetroMAE-distill} respectively, using a fusion score with $\alpha=0.1$.

\section{Further Details: \ourdataset{} Creation} \label{app:further_details_dataset_creation}

\paragraph{License.} \ourdataset{} builds on data from Wikipedia, which carries a Creative Commons Attribution-ShareAlike 4.0 International License. This license requires that any derivative works also carry the same license.

\begin{figure*}[!ht]
    \centering
    \includegraphics[width=0.85\textwidth]{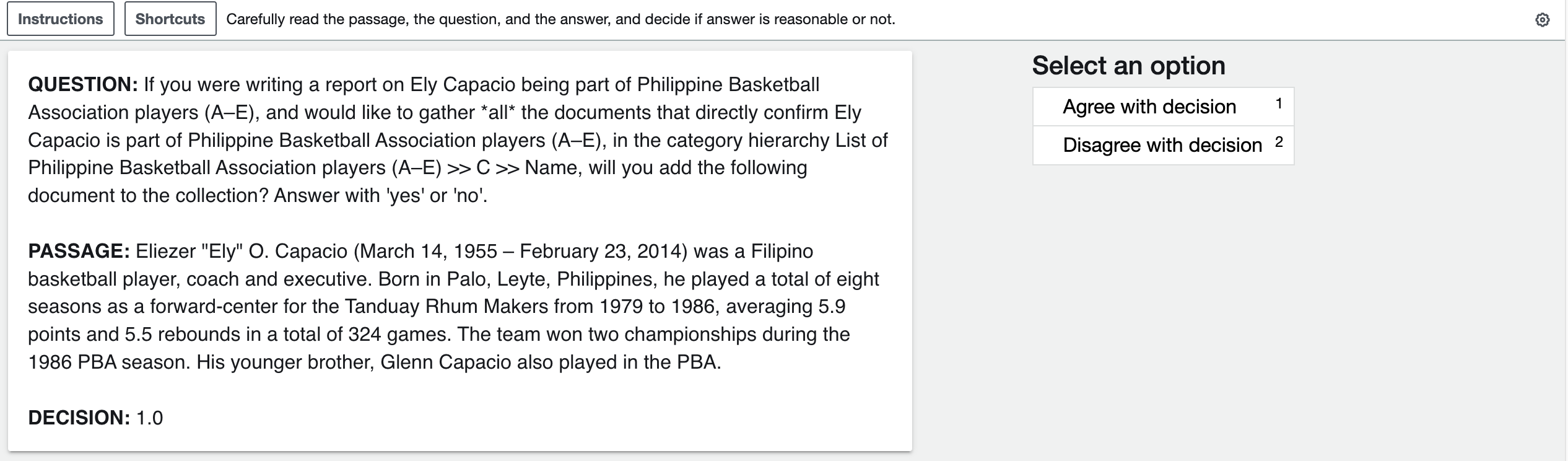}
    \caption{The human evaluation task detailed in \cref{sec:eval_of_dataset_construction}.}
    \label{fig:human_evaluation_task}
\end{figure*}

\paragraph{Conditioning human raters.} Before the evaluation process begins, we need to assure the raters we use understand the task and can perform it adequately. We thus begin a conditioning process. First, we run a qualification exam, and the raters that get all the questions right, are invited to an iterative training process. The process includes small batches, of up to 100 (passage, prompt) pairs, where the rater submits their response and we provide personal feedback. Moreover, all tasks included an option to mark the example as difficult or provide textual feedback about it, to encourage communication from the raters as they work. After each batch raters are filtered out, until we remain with a single rater with a success rate of over 95\% on a single batch. The task is visualized in \cref{fig:human_evaluation_task}.

\paragraph{Automatic identification details.}
\label{app:automatic_filtering}
To automatically identify evidence, \texttt{GPT-4} is provided with a passage and a structured query. In this context, a structured query begins with the article name, followed by its section names arranged hierarchically (separated by ``>>''), corresponding to the structure of the article, and ultimately culminating in the column value. For instance, a typical structured query could be ``Cities and Towns in Cambodia'' (article name) >> ``Cities'' (section name) >> ``Name'' (column name). The task for \texttt{GPT-4} is to determine whether the passage provides evidence supporting the query. The evaluation involves analyzing the text to ascertain whether the passage directly or indirectly confirms the entity in question is part of the group defined by the query. For example, in a query aimed at identifying names of Cambodian cities, the passage must either explicitly state or strongly suggest that a particular city belongs in Cambodia to be considered relevant. Our prompts follow our definition of relevance from \cref{sec:evidence_retrieval_task}:\\
\texttt{If you were writing a report on {member} being part of {article-name}, and would like to gather *all* the documents that directly confirm {member} is part of {article-name}, in the category hierarchy {article-name}  >> {section-name} >> {column-name}, will you add the following document to the collection? Answer with ``yes'' or ``no''.}

\paragraph{Natural-language query generation prompt.}
\label{app:natural_lang_gen}
To translate a structured query to its natural-language variant, we prompt \texttt{GPT-4} using the template below. Examples of input and output can be viewed in \cref{tab:query_examples}.\\
\texttt{Please pretend you are a typical Google Search user, show me what you would write in the search bar. For example: cultural property of national significance in Switzerland:Zurich >> Richterswil >> Name, where >> indicates a hierarchy, a typical search would be: names of cultural properties of national significance in Richterswil, Zurich, Switzerland.}\\ \\
\texttt{Here, try this one: \{input\}} \\

\begin{table}[t]
\centering
\scalebox{0.6}{
\begin{tabular}{>{\centering\arraybackslash}p{6cm}|>{\centering\arraybackslash}p{6cm}}
\textbf{Structured Query} & \textbf{Natural-language Query} \\
\hline
List of Zhejiang University alumni >> Politics \& government >> Name & names of Zhejiang University alumni in politics and government \\
\hline
List of Wisconsin state forests >> Forest name & names of Wisconsin state forests \\
\hline
List of World War I flying aces from the United States >> Served with the Aéronautique Militaire >> Name & names of US World War I flying aces who served with the Aéronautique Militaire \\
\hline
List of LGBT classical composers >> 20th century >> Name & names of 20th century LGBT classical composers \\
\hline
List of Eliteserien players >> Name & names of Eliteserien football players \\
\hline
List of National Monuments in County Sligo >> National Monuments >> Monument name & names of National Monuments in County Sligo, Ireland \\
\end{tabular}}
\caption{Examples of structured queries and their corresponding natural-language form. }
\label{tab:query_examples}
\end{table}

\section{Concordance}\label{app:conc}

\begin{figure*}[htbp]
\centering
    \includegraphics[width=0.4\linewidth]{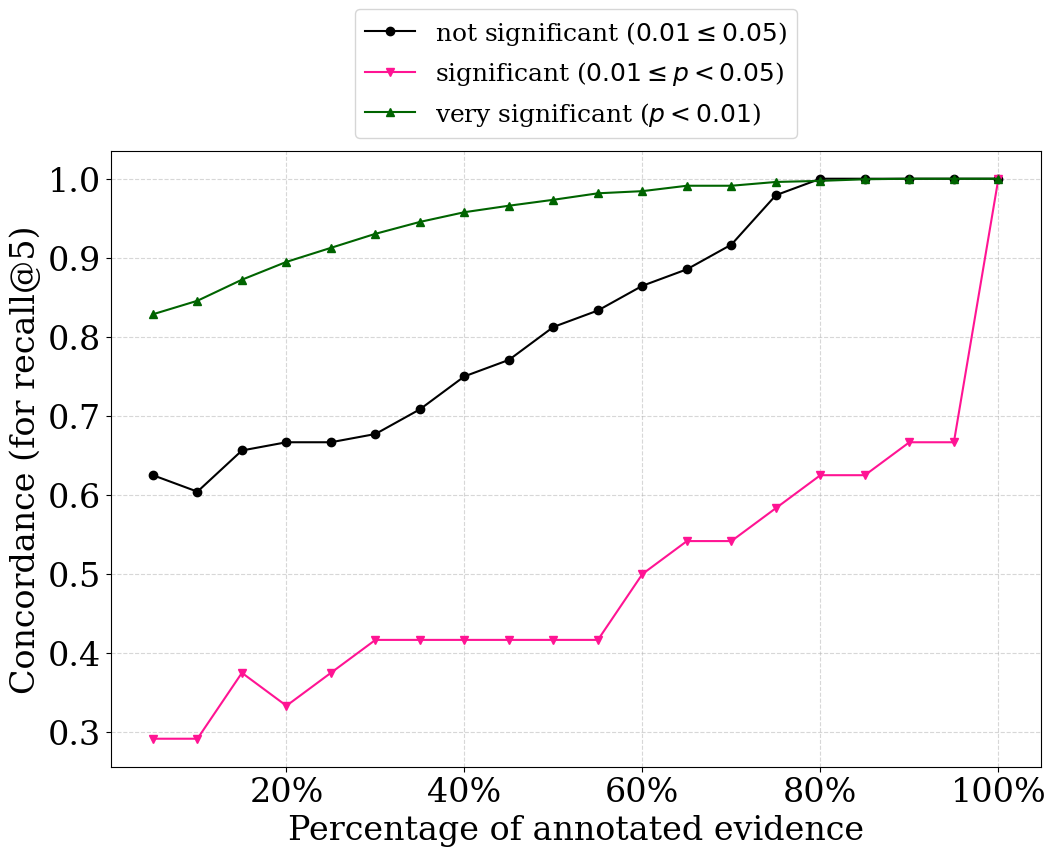}\hfil
    \includegraphics[width=0.4\linewidth]{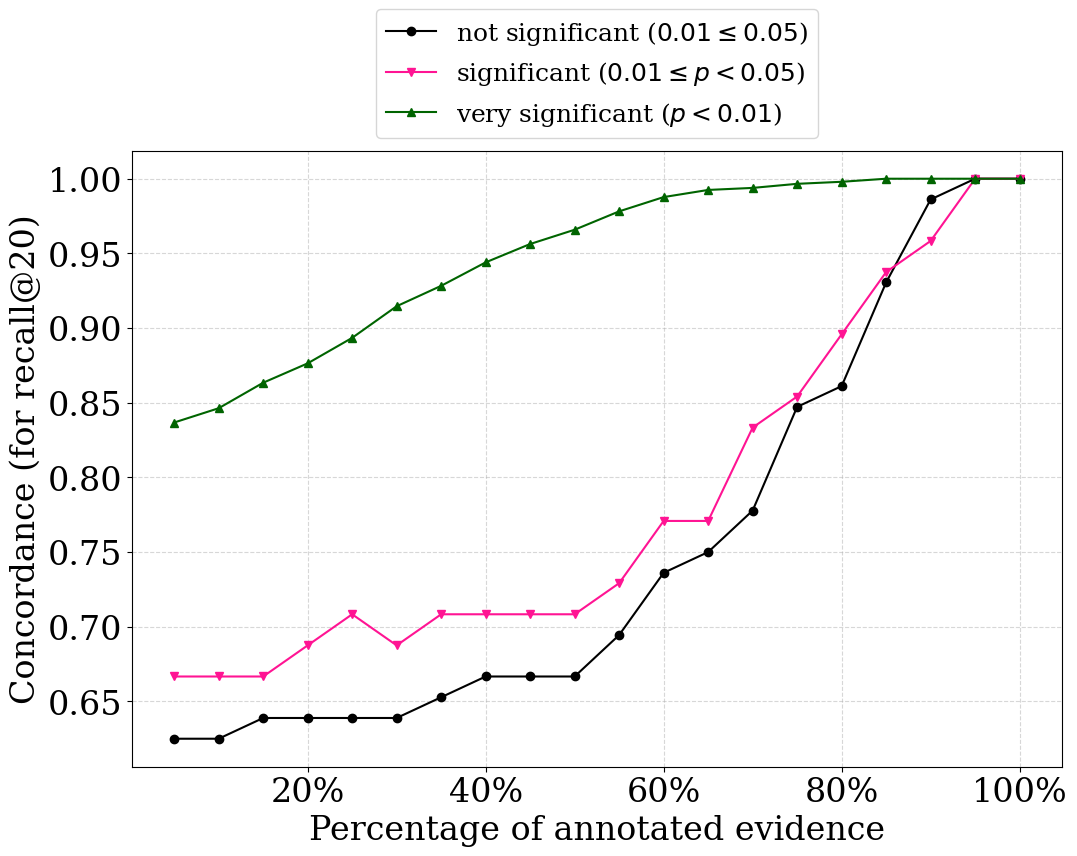}\par\medskip
    \includegraphics[width=0.4\linewidth]{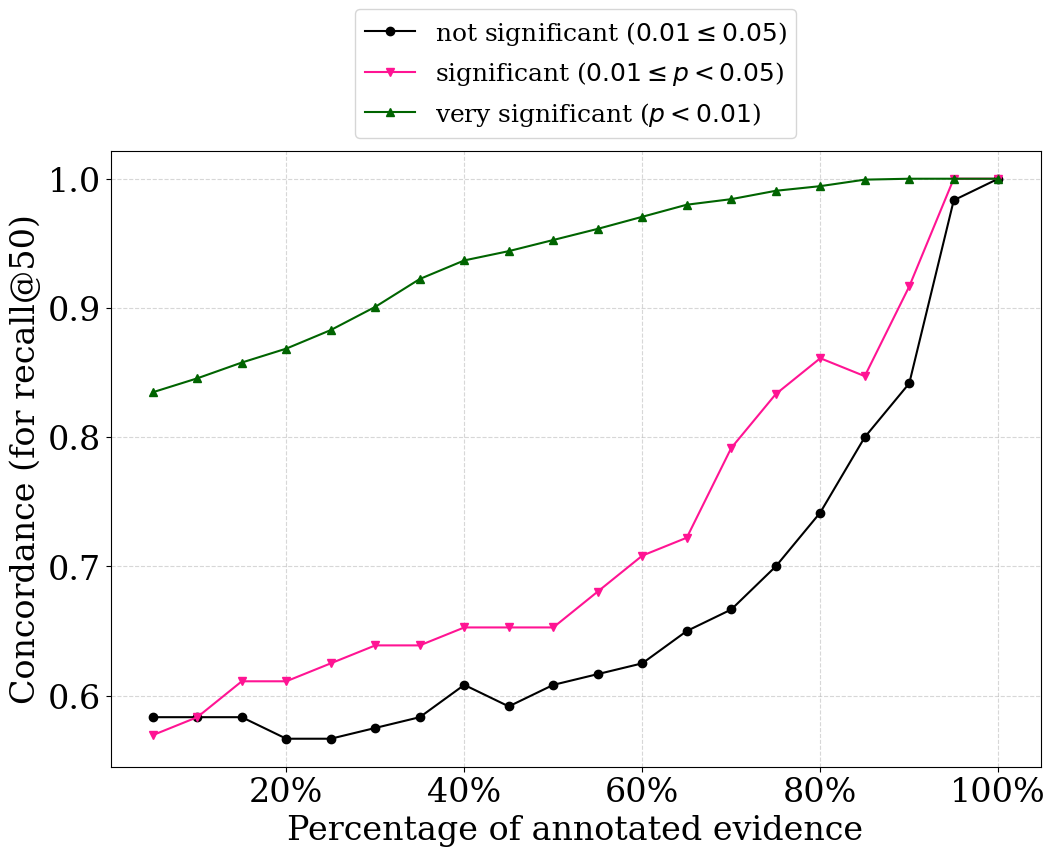}\hfil
    \includegraphics[width=0.4\linewidth]{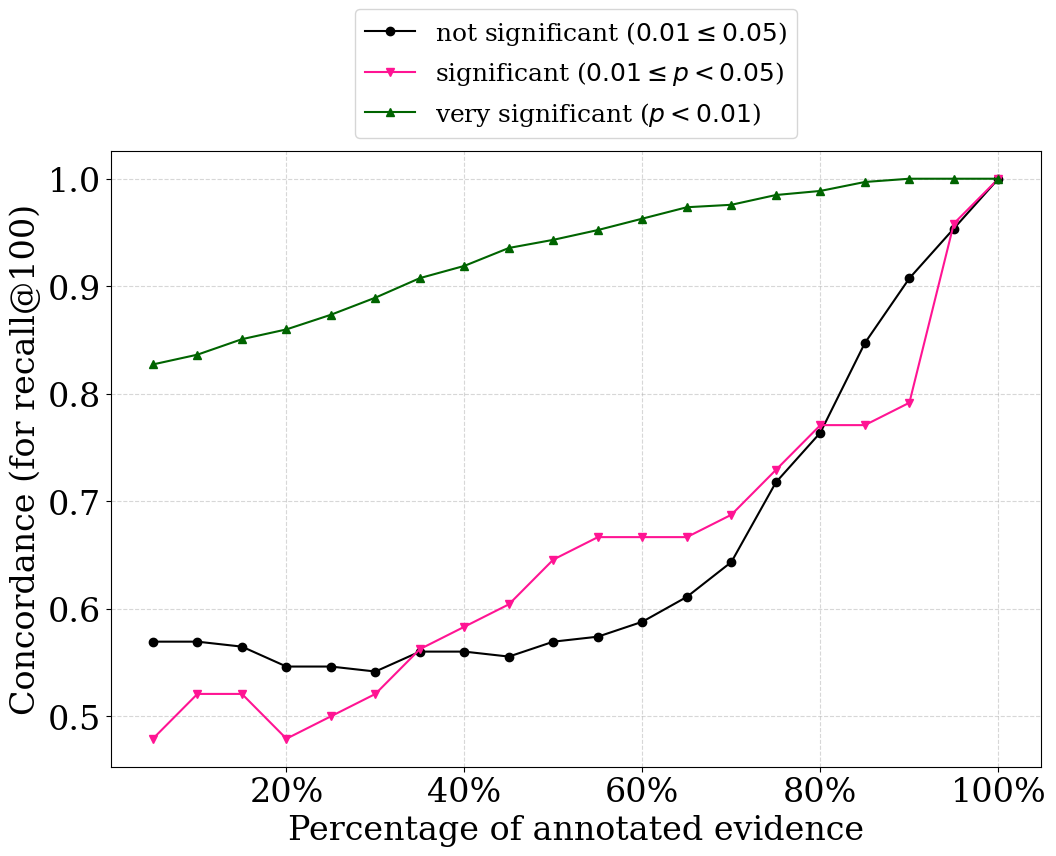}
\caption{Concordance between rankings of systems with varying percentages of evidence and ranking with all evidence, using \recall{5}, \recall{20}, \recall{50}, and \recall{100}. System pairs are divided into 3 buckets as described in \cref{sec:buckets_exp}.}
\label{app:conc_recall}
\end{figure*}
Kendall-$\tau$ \cite{kendall_tau} is a popular metric for evaluating rank correlation between rankings. This is done by comparing the number of concordant and dis-concordant elements between two ranks over a set of elements. More general variants of Kendall-$\tau$ \cite{kendall1945treatment, stuart1953estimation} address cases where ties exist (i.e., in one ranking two elements received an identical score). 

The simplicity of Kendall-$\tau$ makes it tempting to utilize it to compare the ranking of retrieval systems. However, it fails to capture some of the intricacies of this comparison due to several reasons. First, simply comparing system scores is insufficient, as an additional verification using a significance test is necessary. Ties can be defined (i.e., system $A$ is tied with system $B$ if $p>0.05$), but the relation is not transitive ($A$ tied with $B$ and $B$ tied with $C$ does not imply that $A$ is tied with $C$), as required by variants of Kendall-$\tau$ that support ties. Second, some ranking errors are more troublesome than others. Finding that a new system is ``tied'' with the baseline system when in fact it is worse might be undesirable. However, incorrectly reporting that it is better is improper.

Even though Kendall-$\tau$ suffers from the shortcomings above, we hypothesize that it is still a good metric for comparing performance rankings. To validate this we propose a new metric, \emph{concordance}, that addresses these shortcomings of Kendall-$\tau$ and its variants. This is done by considering the relations $A > B$ and $A < B$ for a pair of systems $A$ and $B$. This way if in the ground truth $A$ is significantly better than $B$ and in the compared ranking $A$ is tied with $B$, the two rankings will agree on the relation $A < B$ (will be false in both) and disagree on the relation $A>B$. In a more troublesome error, where $A<B$ in the compared ranking, the two rankings will disagree on both relations.
Formally, let $\pi_1$ and $\pi_2$ be two rankings of a set of retrieval systems $S$. For each pair of systems $s_1, s_2$ and ranking $\pi$ we define
\begin{align*}
    \pi(s_1,s_2) = \begin{cases}
        1, &\text{$s_1$ is significantly better than $s_2$}\\
        0, &\text{otherwise.}
    \end{cases}
\end{align*}
Then concordance is defined as the agreement between the rate of agreement over all ordered pairs of systems between two rankings:
\begin{align*}
    \text{conc}&(\pi_1,\pi_2) = \\ &\frac{1}{P(|S|,2)}\sum_{s_1}\sum_{s_2\neq s_1}\pi_1(s_1,s_2)\odot\pi_2(s_1,s_2),
\end{align*}
where $P(n,r)$ is the number of permutations of size $r$ from a set of size $n$, and $\odot$ is the XNOR operator (equals to $1$ if both inputs equal).

Using concordance, we validate the results found in \cref{sec:buckets_exp} and \cref{sec:effect_of_number_of_relevant_passages} using Kendall-$\tau$. This is done by repeating the experiment and calculating the mean concordance of system rankings given evidence found by different systems with the ground truth ranking (in which all evidence are annotated). We run this experiment for a single annotated evidence and different percentiles of annotated evidence. 

In Table~\ref{tab:conc_single_relevant_buckets_all_k} and ~\cref{app:conc_recall} we see that pairs of systems with a very significant difference between them (i.e., $p<0.01$) are evaluated with higher accuracy than systems falling in the other two buckets. This validates the results found in \cref{sec:buckets_exp} and \cref{sec:effect_of_number_of_relevant_passages} and shows that Kendall-$\tau$ is a good proxy for evaluating the rankings of IR systems. 

\begin{table}[h]
\centering
\scalebox{0.75}{
\begin{tabular}{c|cc|c}
\textbf{k}&\textbf{$p_{min}$} & \textbf{$p_{max}$} & \textbf{Concordance} \\ 
\hline

5 & 0.0 & 0.01 & 0.809   \\
5 & 0.01 & 0.05 & 0.292  \\
5 & 0.05 & 1.0 & 0.646  \\ \hline\hline

20 & 0.0 & 0.01 & 0.823   \\
20 & 0.01 & 0.05 & 0.708  \\
20 & 0.05 & 1.0 & 0.611  \\ \hline\hline

50 & 0.0 & 0.01 & 0.821   \\
50 & 0.01 & 0.05 & 0.556  \\
50 & 0.05 & 1.0 & 0.592  \\ \hline\hline

100 & 0.0 & 0.01 & 0.813   \\
100 & 0.01 & 0.05 & 0.500 \\
100 & 0.05 & 1 & 0.583 

\end{tabular}}
\caption{Concordance computed only on pairs of systems that fall within the [$p_{min}$, $p_{max}$) bucket. k is the \recall{k} used.}
\label{tab:conc_single_relevant_buckets_all_k}
\end{table}

\section{TREC Coverage} \label{app:trec_coverage_simulation}

TREC \cite{trec2019,trec2020,trec2021,trec2022, trec2023}, a popular retrieval competition, also tries to deal with the problem of partial annotated retrieval datasets.
In this section we compare our approach for collecting multiple evidence for queries with their approach. This is done by applying TREC's approach to our dataset and testing its coverage. This will reveal, even though anecdotally, the ability of TREC's approach to find numerous evidence. The approach in TREC does not utilize a structured data source for the creation of the judgement set. Instead, they create a pool of candidates from the set of passages retrieved by a large set of systems. Specifically, TREC runs a competition and publishes a query set and a corpus. Any participant team executes their system and submits a retrieved list. Then, TREC pools top-$k$ passages from each participant and sends them for human annotation, annotating for relevancy. Before applying the approach used by TREC to our dataset we first formally define this process. Let $Q$ be the set of queries and $E_q$ the evidence set of query $q\in Q$. In addition, let $S$ be the set of systems and $E_{q,s}$ be the evidence set found in the top-$10$ passages retrieved by system $s\in S$ for query $q\in Q$. Then, the judgement set of query $q$ is defined as $J_q(S)=\cup_{s\in S}E_{q,s}$. We denote the coverage of $S$ on $Q$ as:

$$ C_Q(S) = \frac{1}{|Q|}\sum_{q\in Q}\frac{|J_q(S)|}{|E_q|}.$$

When fixing the number of passages retrieved by each system to $k=10$, as done in TREC, and given the $12$ systems considered in this paper (see Section~\ref{sec:experiment_setup}), we can compute their coverage on \ourdataset{} which is equal to $31.7\%$. While this may be low, we only consider a small number of systems, as it is typical to use around $100$ systems. Also, increasing $k$ is expected to increase the coverage. Following, we use extrapolation techniques to estimate the affect of both.

\subsection{Extrapolating Number of Systems}

\begin{figure}[ht!]
    \includegraphics[width=7cm]{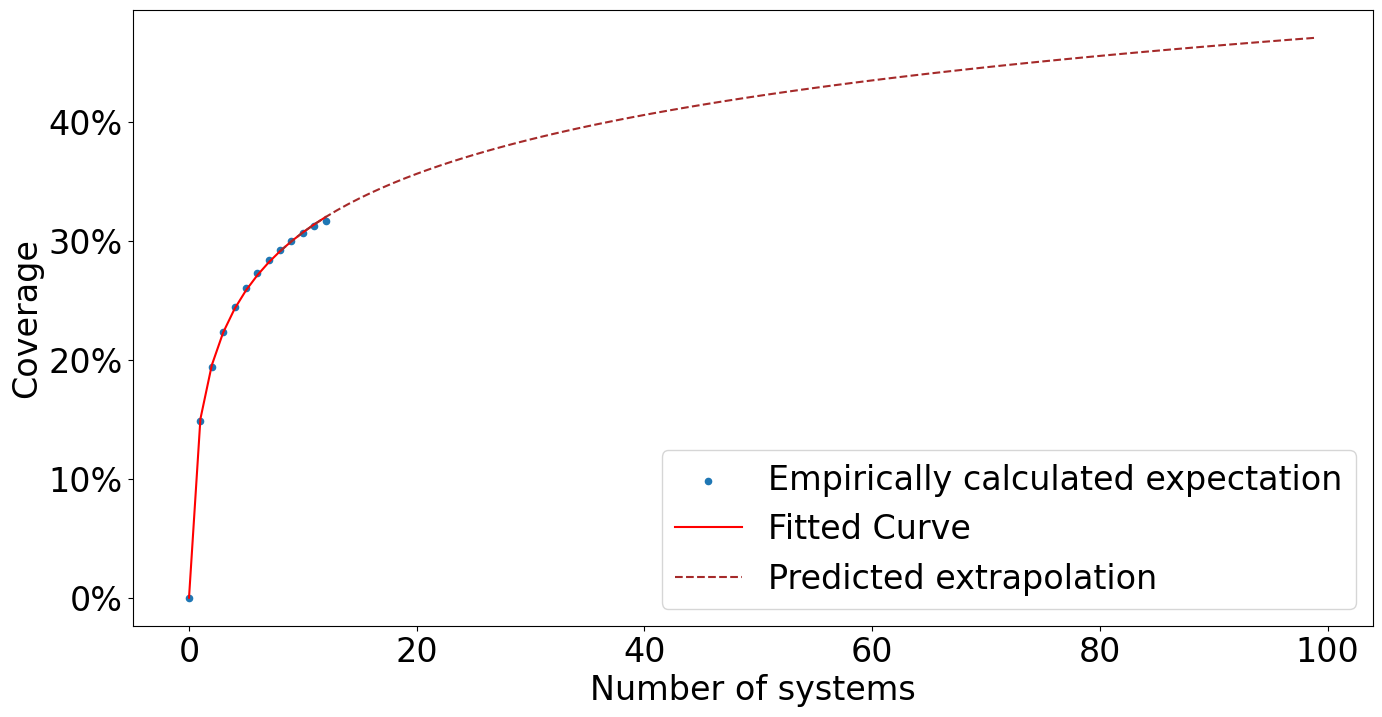}
    {\caption{Fraction of relevant passages covered by top-10 passages for $s$ systems.}\label{fig:trec-numsys}}
\end{figure}

Due to time and compute constraints using $100$ systems, as typically done in the TREC competition, is unrealistic. This leads us to approximate the coverage instead. In order to approximate the coverage of a larger number of systems we first fix $k=10$, and compute the expected coverage of a random subset of systems of size $t$ uniformly sampled from $S$. That is,

$$ C^*_Q(S,t) = \mathop{\mathbb{E}}_{S'\sim U(S),~|S'|=t}[C_Q(S')]. $$
Given the values of $C^*_Q(S,t)$ for $t=1,\ldots, 12$, we fit a logarithmic curve (as coverage is both concave and monotonically-increasing) to these observations and observe a root mean-squared-error (RMSE) of $0.16\%$ and a maximum error of $0.31\%$. Finally we extrapolate to predict the coverage for $t=13,\ldots, 100$. The results of the experiment is presented in \cref{fig:trec-numsys}. As can be seen, we predict that broadening the judgement sets by retrieving with as many as $100$ systems only increases the coverage from $31.7\%$ to $47.1\%$. This result further corroborates the finding by \citep{how_reliable_are_the_results_of_large_scale_info_retrieval_experiments}, which states that the pooling approach used in TREC finds, at best, 50-70\% of the evidence. We conclude that our approach is able to achieve a much higher coverage. This is expected to improve the correctness of our evaluation. Note that our approach depends on structured data in Wikipedia. On the other hand, the approach utilized in TREC is universal as it can be applied to any corpus and query.

\subsection{Extrapolating Number of Retrieved Documents per System}
Increasing the pool size can uncover additional positive results, but will result in a significantly larger annotation pool size. We adopt a similar method to extrapolating the coverage by increasing the number of systems, and but focus instead on the size of the pool.

We use the coverage evaluation dataset described in section~\ref{sec:eval_of_dataset_construction} which takes a the top-20 pool from 12 systems and uses human annotators to label the relevancy of each entry in the pool. Next, we assign each relevant entry in the pool its minimum rank from all systems and construct pools for each depth size. For example, for k=10, we take all documents that were ranked at the top-10 by at least a single system.

\begin{figure}[ht!]
    \includegraphics[width=7cm]{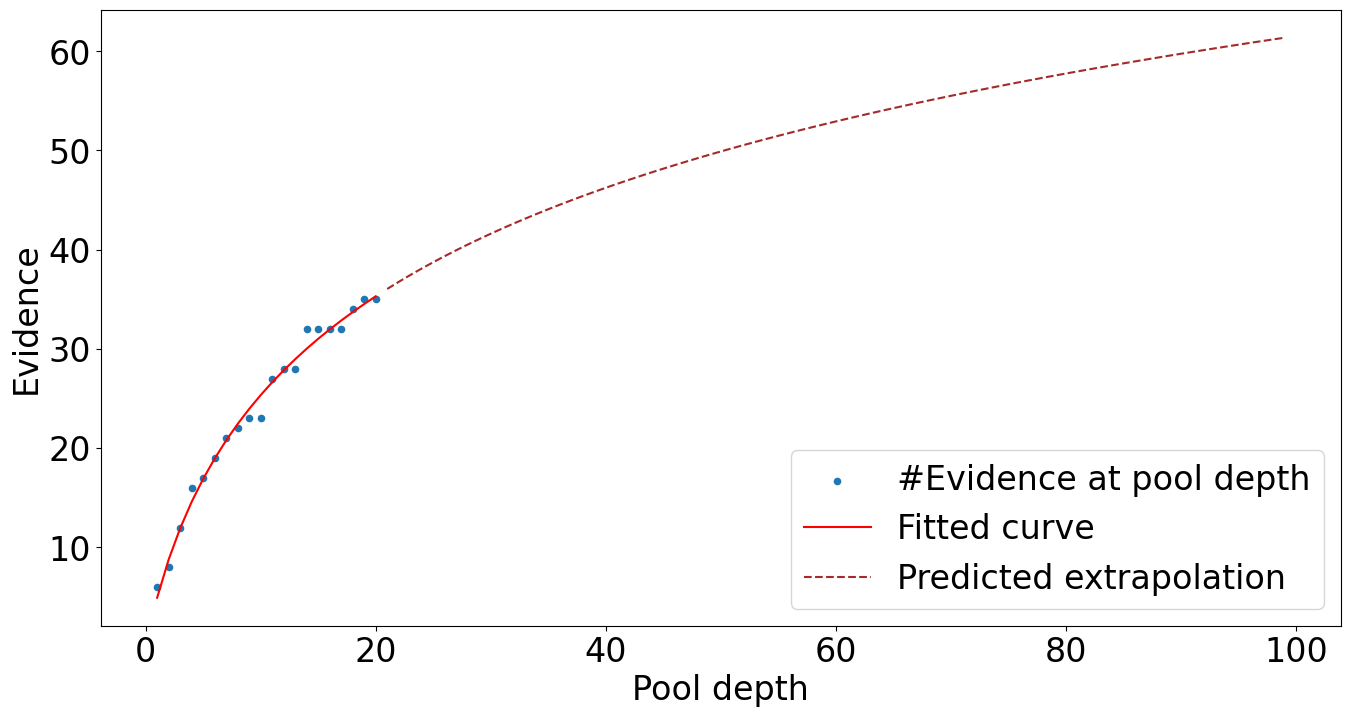}
    {\caption{Number of newly identified evidence by pool depth $k$.}
    \label{fig:trec-newpool}}
\end{figure}

\begin{figure}[ht!]
    \includegraphics[width=7cm]{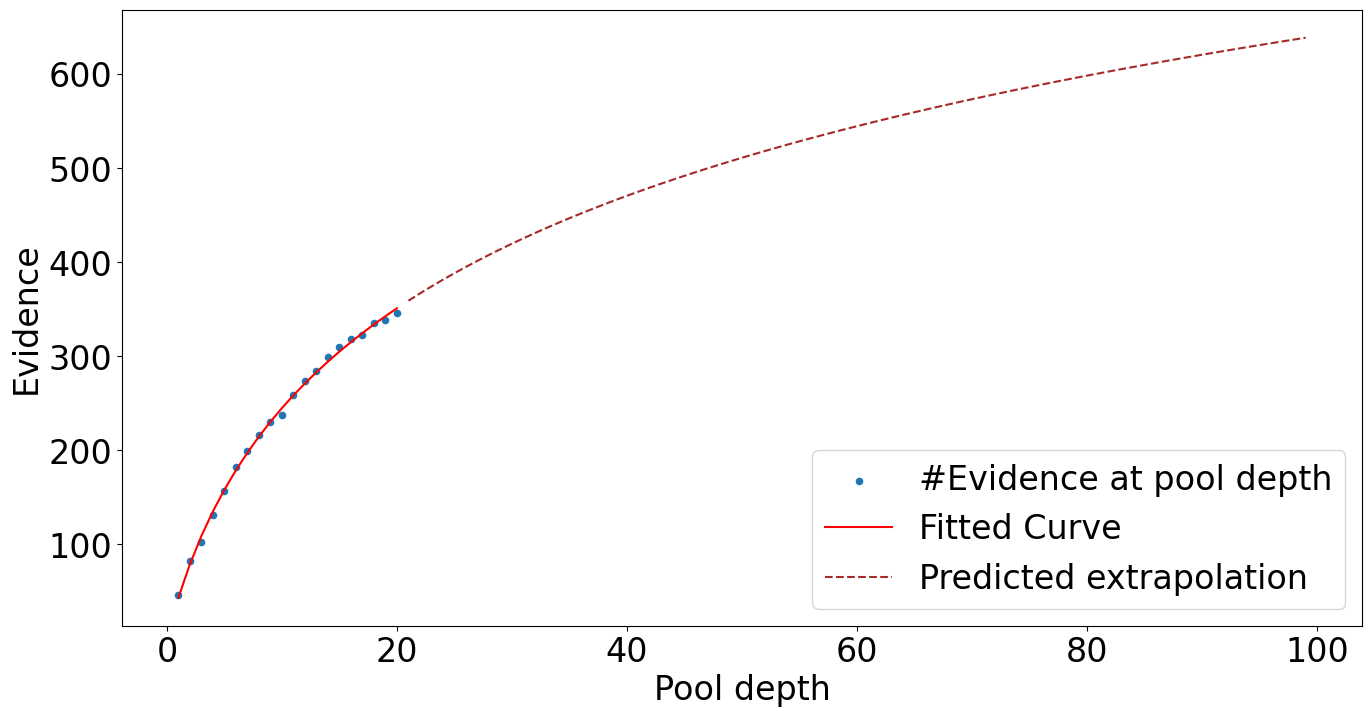}
    {\caption{Number of identified evidence by pool depth.}
    \label{fig:trec-numpool}}
\end{figure}

Finally, we extrapolate to predict for the number of newly identified evidence (~\cref{fig:trec-newpool}) and the overall documents found by the pooling approach (~\cref{fig:trec-numpool}) for $t=21,\ldots, 100$. The results show that even for a pool-depth of $k=100$, we estimate that only 60 new evidences will be identified. This means that the coverage of our method is estimated to be $\sim94.5\%$ out of all identified evidence. In addition, we see that the pooling approach for $k=100$ is estimated to retrieve 638 evidence (578 already found by our method) covering only $60.8\%$ with a significant increase of annotation overhead.

\clearpage

\section{Extended Results} \label{app:more_results}
In the main paper we focused on \recall{20} for brevity when reporting results. 
Here, we report experiments shown in \cref{sec:exp_study} measuring also \recall{5/50/100}.
Conclusions pointed out in the main paper hold for all values of $k$.







\begin{table}[h]
\centering
\scalebox{0.7}{
\begin{tabular}{c|cc|c|c}
\textbf{k}&\textbf{$p_{min}$} & \textbf{$p_{max}$} & \textbf{partial-$\tau$}& \textbf{Error-rate (\%)} \\ 
\hline

5 & 0.0 & 0.01 & 0.654 & 17.30\  \\
5 & 0.01 & 0.05 & -0.583 & 79.15\ \\
5 & 0.05 & 1.0 & -0.125 & 56.25 \ \\ \hline\hline

20 & 0.0 & 0.01 & 0.658 & 17.10\  \\
20 & 0.01 & 0.05 & 0.333 & 33.35\ \\
20 & 0.05 & 1.0 & 0.000 & 50.00\ \\ \hline\hline

50 & 0.0 & 0.01 & 0.658 & 17.10\  \\
50 & 0.01 & 0.05 & 0.167 & 41.65\ \\
50 & 0.05 & 1.0 & 0.200 & 40.00\ \\ \hline\hline

100 & 0.0 & 0.01 & 0.642 & 17.90\  \\
100 & 0.01 & 0.05 & -0.083 & 54.15\ \\
100 & 0.05 & 1 & 0.185 & 40.75\

\end{tabular}}
\caption{partial-Kendall-$\tau$ similarity (as defined in \cref{sec:buckets_exp}, denoted here as partial-$\tau$) and Error-rate computed only on pairs of systems that fall within the [$p_{min}$, $p_{max}$) bucket. k is the \recall{k} used.}
\label{tab:kendall_tau_single_relevant_buckets_all_k}
\end{table}







\begin{table}[h]
\centering
\scalebox{0.7}{
\begin{tabular}{c|c|c|c}
\textbf{k} & \textbf{Selection} & \textbf{$\tau$-similarity}& \textbf{Error-rate (\%)} \\ 
\hline

5 & Random & 0.815 & \hspace{0.13cm} 9.25\ \\ \hline
5 & Most popular      & 0.727 &  13.65\ \\
5 & Longest      & 0.462 & 26.90\ \\
5 & Shortest      & 0.585 & 20.75\ \\ \hline
5 & System-based & 0.587 & 80.65\ \\ \hline\hline

20 & Random & 0.936 & \hspace{0.2cm} 3.20 \ \\ \hline
20 & Most popular      & 0.697 &  15.15\ \\
20 & Longest      & 0.545 & 22.75\ \\
20 & Shortest      & 0.697 & 15.15\ \\ \hline
20 & System-based & 0.616 & 19.20\ \\ \hline\hline

50 & Random & 0.916 & \hspace{0.13cm}  4.20\ \\ \hline
50 & Most popular      & 0.687 &  15.65\ \\
50 & Longest      & 0.606 & 19.70\ \\
50 & Shortest      & 0.576 & 21.20\ \\ \hline
50 & System-based & 0.596 & 20.20\ \\ \hline\hline

100 & Random & 0.894 & \hspace{0.13cm}  5.30\ \\ \hline
100 & Most popular      & 0.818 & \hspace{0.13cm} 9.10\ \\
100 & Longest      &  0.697&15.15\ \\
100 & Shortest      &  0.545& 22.75\ \\ \hline
100 & System-based & 0.523 & 23.85\ \\

\end{tabular}}
\caption{Kendall-$\tau$ similarities and error for different biases, in a single-annotation setup. k is the \recall{k}.}
\label{tab:kendall_tau_single_relevant_all}
\end{table}

\begin{figure*}
    \centering
    \begin{minipage}{.32\textwidth}
        \includegraphics[width=\linewidth]{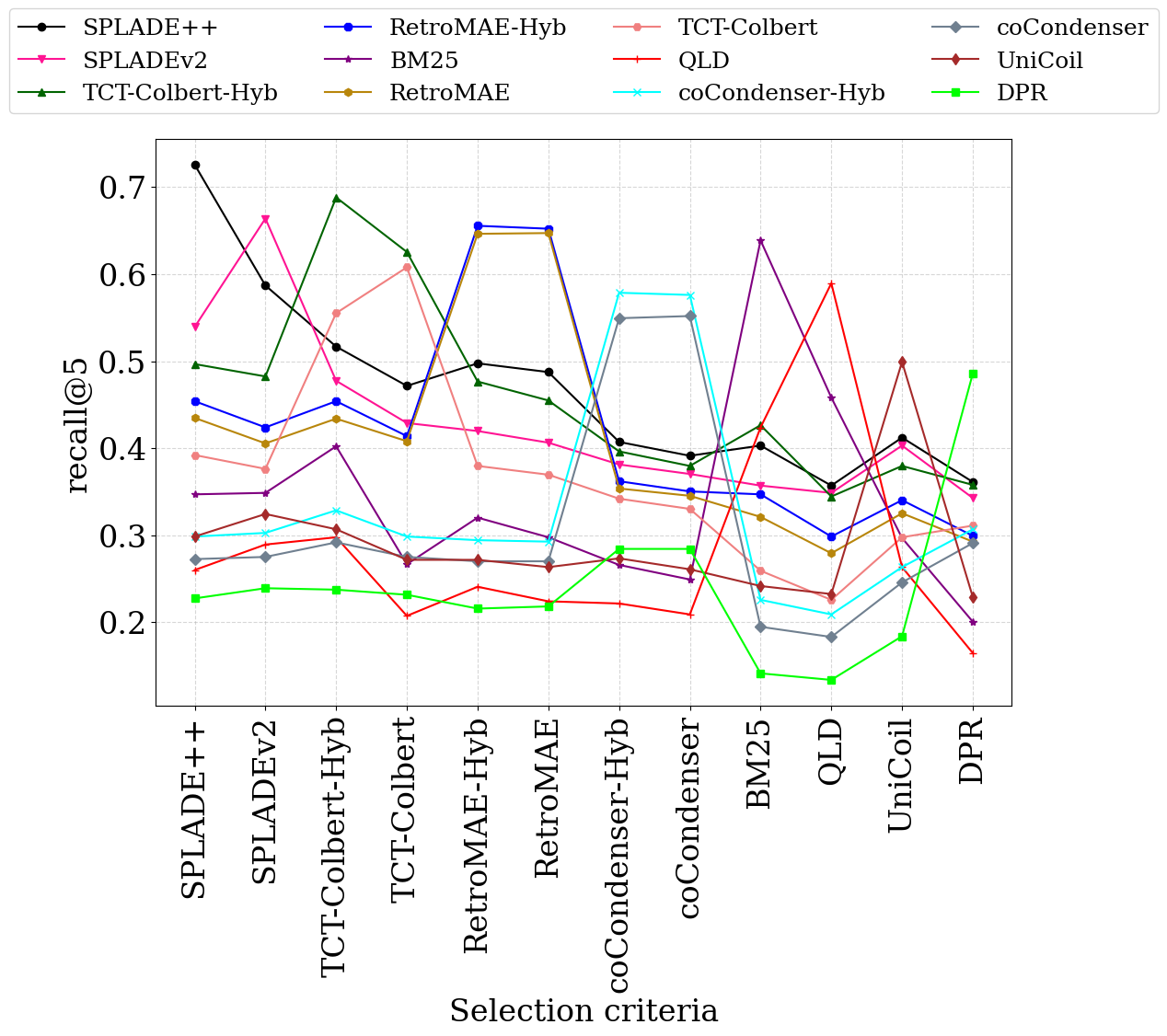}
    \end{minipage}\hfill
    \begin{minipage}{.32\textwidth}
        \includegraphics[width=\linewidth]{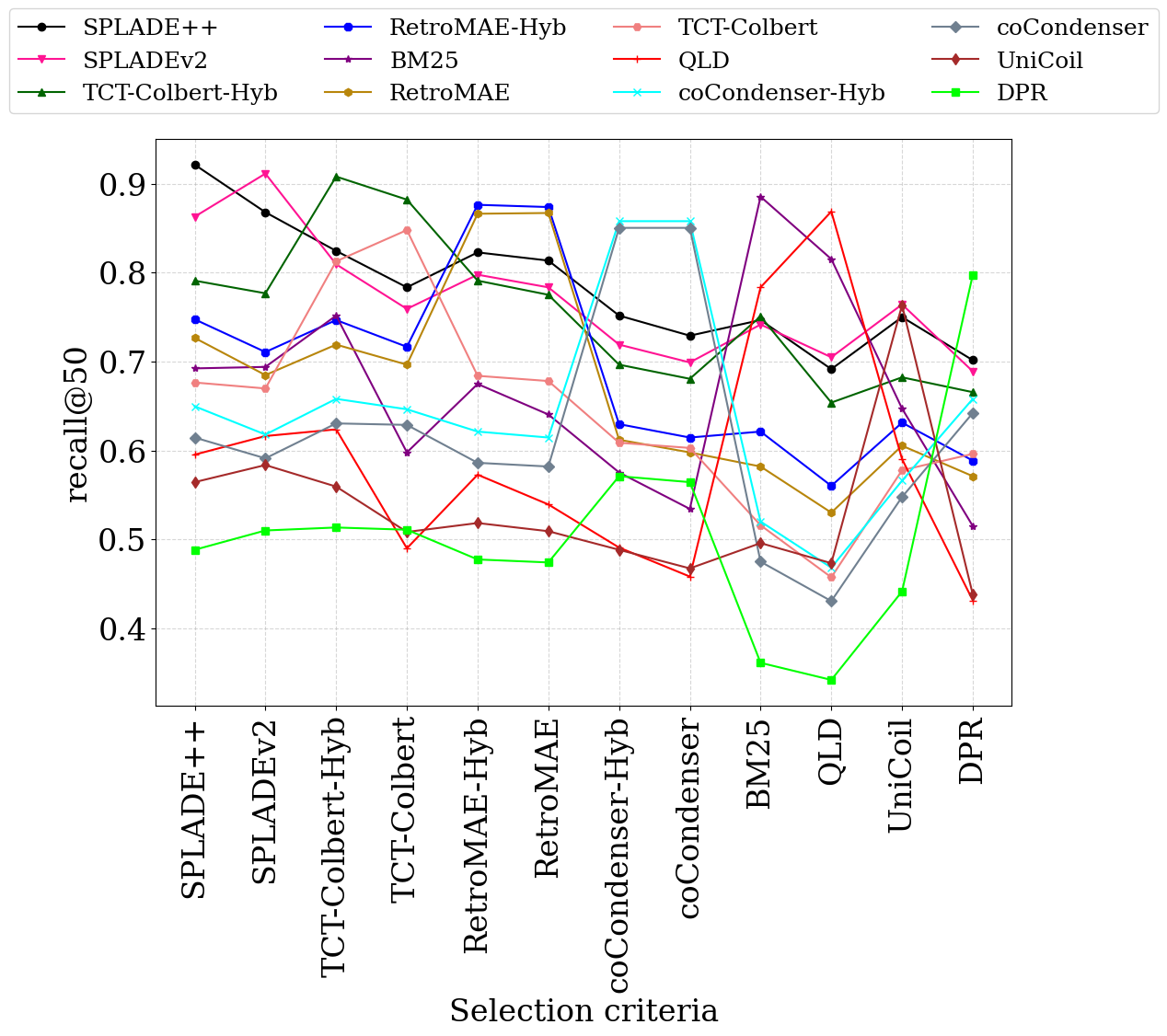}
    \end{minipage}\hfill
    \begin{minipage}{.32\textwidth}
        \includegraphics[width=\linewidth]{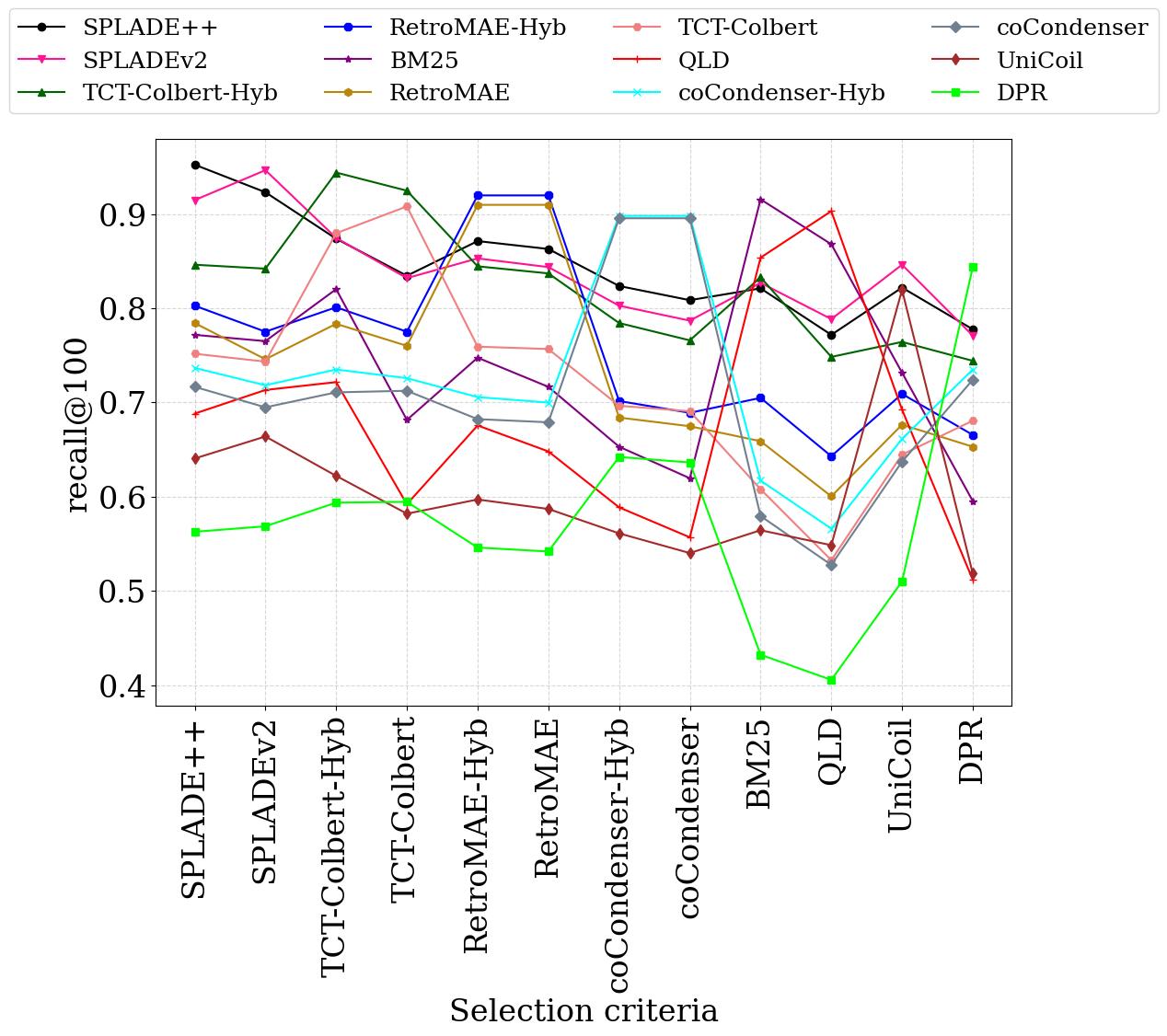}
    \end{minipage}
        
    \caption{Single-annotation per query datasets with varying selection methods. Left to right: \recall{5/50/100}.}
    \label{app:swaps}
\end{figure*}

\begin{figure*}
    \centering
    \begin{minipage}{.32\textwidth}
        \includegraphics[width=\linewidth]{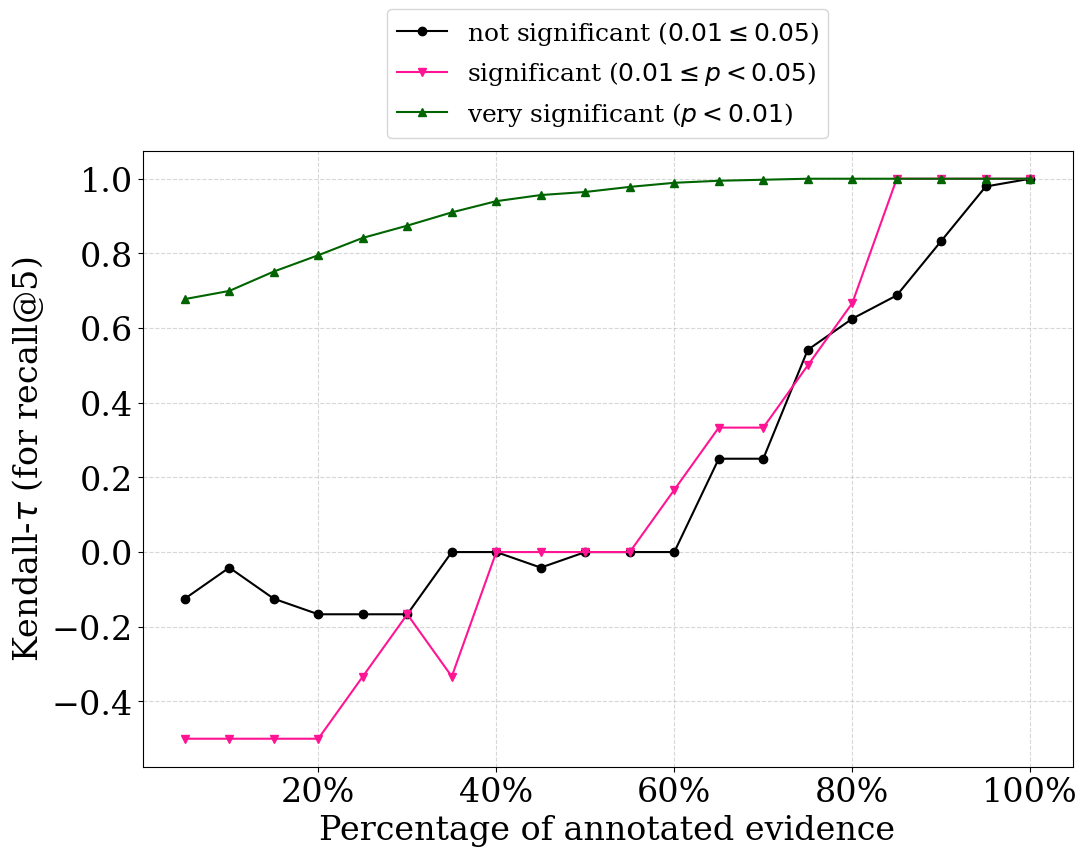}
    \end{minipage}\hfill
    \begin{minipage}{.32\textwidth}
        \includegraphics[width=\linewidth]{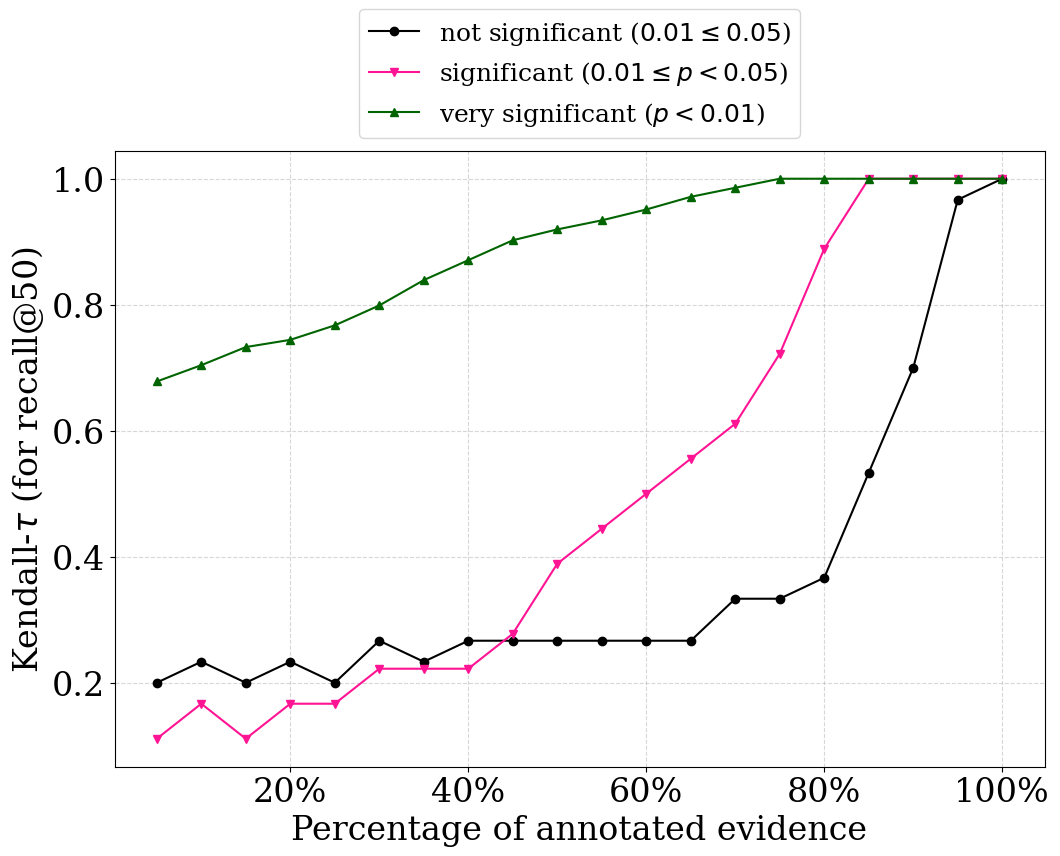}
    \end{minipage}\hfill
    \begin{minipage}{.32\textwidth}
        \includegraphics[width=\linewidth]{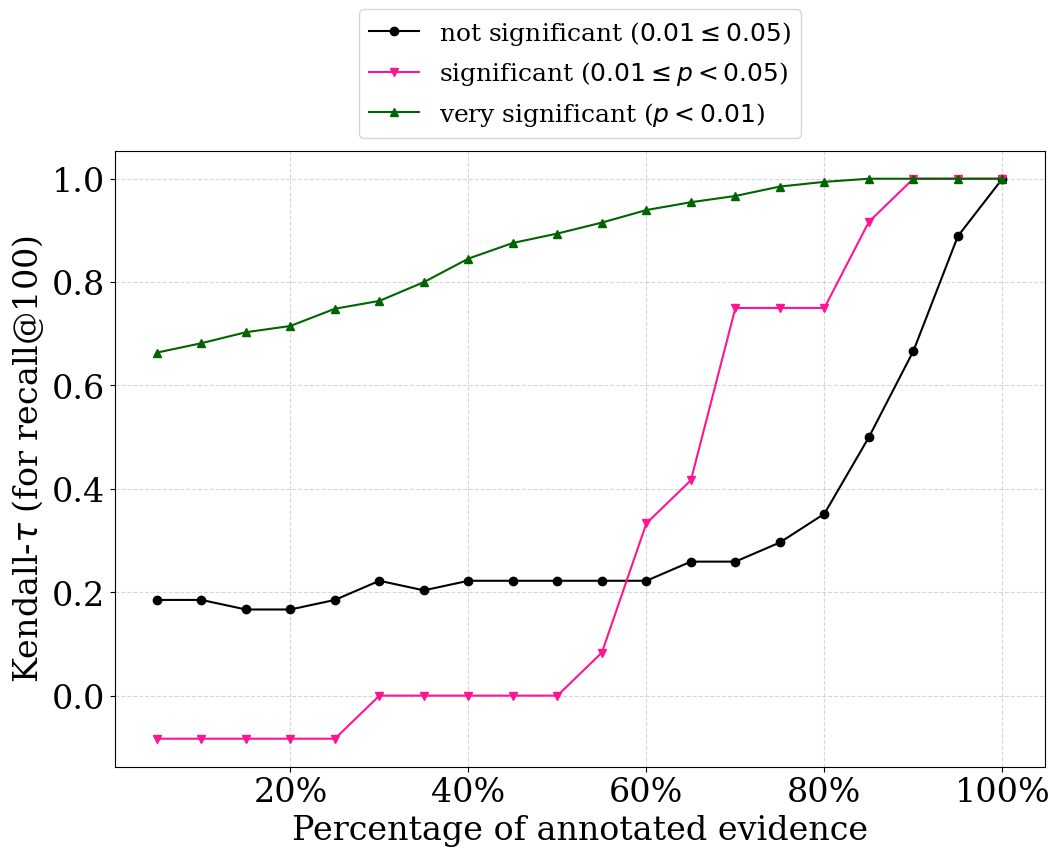}
    \end{minipage}
        
    \caption{Kendall-$\tau$ between rankings of systems with varying percentages of evidence and ranking with all evidence, using \recall{5/50/100}. System pairs are divided into 3 buckets as described in \cref{sec:buckets_exp}.}
    \label{app:kendall_tau_recall}
\end{figure*}



\begin{figure*}[!ht]
    \centering
    \scalebox{0.85}{
        \includegraphics[width=0.75\textwidth]{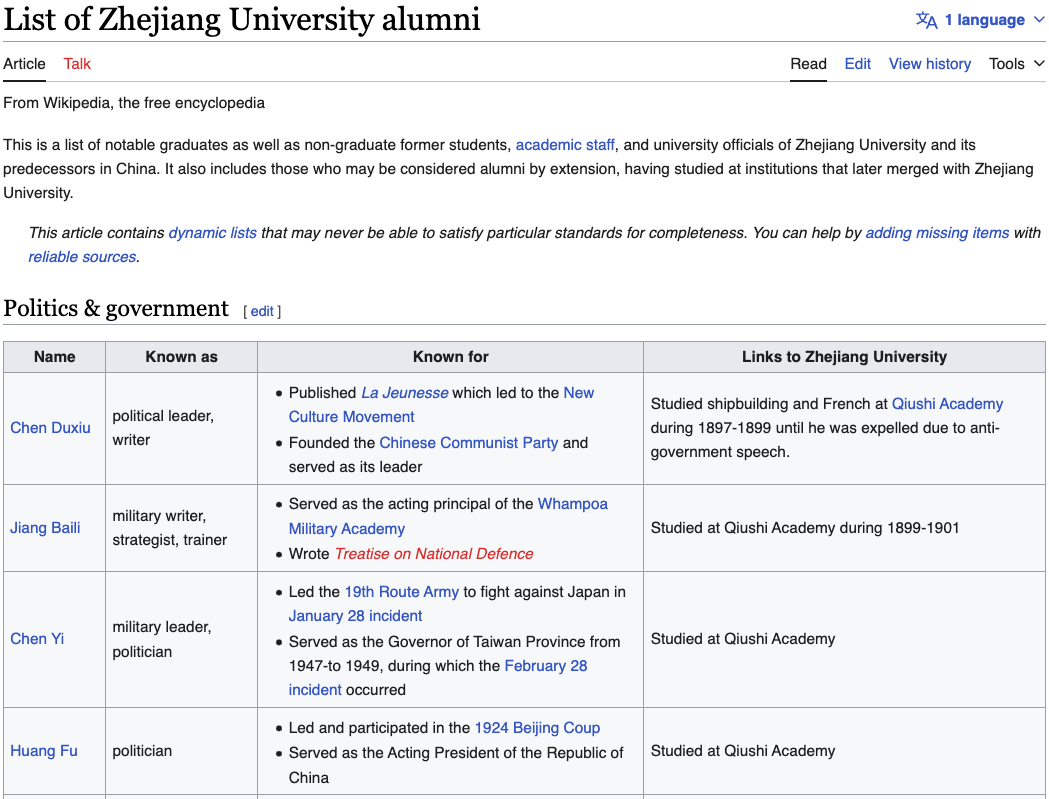}
    }
    \caption{A screenshot of the Wikipedia article corresponding to the first query in \cref{tab:query_examples}.}
    \label{app:example_of_list_of_page}
\end{figure*}

\end{document}